\documentclass[pdflatex,prd,twocolumn,showpacs,superscriptaddress,nofootinbib]{revtex4-1}
\usepackage{bm,amsmath,amssymb}
\usepackage{graphicx}
\usepackage{subfigure}
\usepackage{color}
\usepackage{xcolor}
\usepackage{soul}
\usepackage{hyperref}
\usepackage{multirow}
\usepackage{lineno}
\usepackage{lipsum}
\usepackage{physics}
\hypersetup{
    colorlinks = false,
    urlcolor = red
    }

\newcommand \beqa {\begin{eqnarray}}
\newcommand \eeqa {\end{eqnarray}}

\def\lsim{\raise0.3ex\hbox{$<$\kern-0.75em\raise-1.1ex\hbox{$\sim$}}}
\def\gsim{\raise0.3ex\hbox{$>$\kern-0.75em\raise-1.1ex\hbox{$\sim$}}}

\usepackage{graphicx}

\usepackage{fancyhdr}
\usepackage{etoolbox}  

\graphicspath{{figures2/}}

\begin{document}


\title{Generalized susceptibilities and the properties of charm degrees of freedom across the QCD crossover temperature}


\author{Olaf Kaczmarek}
\affiliation{Fakult\"at f\"ur Physik, Universit\"at Bielefeld, D-33615 Bielefeld, Germany}
\author{Frithjof Karsch}
\affiliation{Fakult\"at f\"ur Physik, Universit\"at Bielefeld, D-33615 Bielefeld, Germany}
\author{Peter Petreczky}
\affiliation{Physics Department, Brookhaven National Laboratory, Upton, New York 11973, USA}
\author{Christian Schmidt}
\affiliation{Fakult\"at f\"ur Physik, Universit\"at Bielefeld, D-33615 Bielefeld, Germany}
\author{Sipaz Sharma}
\affiliation{Physik Department, Technische Universit\"at M\"unchen, James-Franck-Stra{\ss}e 1, D-85748 Garching~b.~M\"unchen, Germany}
\collaboration{HotQCD collaboration}
\date{\today}

\begin{abstract}

We study the generalized charm susceptibilities in 2+1 flavor QCD on the lattice at several lattice spacings. We show that, below the chiral crossover, these susceptibilities are well described by the hadron resonance gas (HRG) model if charmed hadrons not listed in tables of the Particle Data Group are included. However, the HRG description abruptly breaks down just above the chiral crossover. To understand this, we use a model for the charm pressure in which it is expressed as the sum of partial pressures from charmed baryons, charmed mesons, and charm quarks. We present continuum estimates of these partial pressures and find that, while the partial pressures of charmed mesons and baryons drop below their respective HRG predictions, the charm quark pressure becomes non-zero above the chiral crossover.

\end{abstract}

\pacs{11.10.Wx, 11.15.Ha, 12.38.Aw, 12.38.Gc, 12.38.Mh, 24.60.Ky, 25.75.Gz, 25.75.Nq}

\maketitle
\thispagestyle{fancy}
\fancyhf{} 
\fancyhead[R]{TUM-EFT 197/25} 
\renewcommand{\headrulewidth}{0pt} 

\section{Introduction}
Strong interaction matter undergoes a transition at $T_{pc}=(156.5\pm 1.5)$ MeV to a new state characterized by restoration of the chiral symmetry
\cite{HotQCD:2018pds}. This state of matter is studied experimentally
in heavy ion collisions, see Refs.~\cite{Busza:2018rrf,Harris:2023tti} for recent
reviews.
Owing to their large masses, charm and bottom quarks are produced only during the earliest stages of a heavy-ion collision.
They are affected by the hot-dense medium during its entire evolution, and ultimately hadronize into open and hidden heavy-flavor hadrons.  
From the yields of these hadrons, it is possible to gain valuable insights into the nature of the produced medium, in particular
the strong collective behavior of the produced medium, where
even heavy quarks/hadrons are involved in the collective behavior. It has been suggested early on that the heavy-flavor hadrons do provide an excellent
probe for the deconfining nature of the quark-hadron transition in strong interaction matter \cite{Matsui:1986dk}. This led to
a large experimental and phenomenological program
\cite{He:2022ywp,Rapp:2018qla} dedicated to the 
study of properties of charmed hadrons in the QCD
transition region \cite{Braun-Munzinger:2000csl,Braun-Munzinger:2000eyl,Andronic:2017pug}.
To interpret the results of the experiments, however, it is important to know the nature of the charm and bottom degrees of freedom across the QCD transition. 

In order to study the nature of the charm degrees of freedom, one
can calculate in the framework of lattice QCD, the 
so-called generalized charm susceptibilities, {\it i.e.} derivatives of the QCD pressure ($P$),
\begin{equation}
    P=\frac{T}{V}\ln Z(T,V,\hat{\mu}_B,\hat{\mu}_Q,\hat{\mu}_S,\hat{\mu}_C)
    \; ,
\label{pressure}
\end{equation}
with respect to chemical potentials coupling to the conserved currents of net baryon number ($B$), electric charge ($Q$), strangeness ($S$)
and charm ($C$),
\begin{equation} 
{\chi^{BQSC}_{klmn}=\dfrac{\partial^{(k+l+m+n)}\;\;[P\;(\hat{\mu}_B,\hat{\mu}_Q,\hat{\mu}_S,\hat{\mu}_C)\;/T^4]}{\partial\hat{\mu}^{k}_B\;\;\partial\hat{\mu}^{l}_Q\;\;\partial\hat{\mu}^{m}_S\;\;\partial\hat{\mu}^{n}_C}}\bigg|_{\vec{\mu}=0} \; .
\label{eq:chi}
\end{equation}
Here we introduced a dimensionless notation for chemical potentials, 
${\hat{\mu}_X = \mu_X/T}$, with $X \in \{B, Q, S, C\}$.
The first study of generalized charm susceptibilities has been performed about a decade ago \cite{Bazavov:2014yba,Mukherjee:2015mxc}. These
calculations have been performed on rather coarser lattices
corresponding to temporal extent $N_{\tau}=6$ and 8. We recall
that the lattice spacing, $a$, is related to the temperature, $T$, and 
temporal extent, $N_\tau$, as $a=1/(T N_{\tau})$. The statistics was also
very limited in the above studies. Despite these limitations, the study of generalized charm susceptibilities provided important insights 
into the nature of the charm degrees of freedom at different
temperatures. It showed that below the chiral crossover temperature,
the generalized charm susceptibilities are well described by a
gas of uncorrelated open-charm hadrons, while new degrees of freedom
appear above the chiral crossover temperature \cite{Bazavov:2014yba,Mukherjee:2015mxc}. 

Thermodynamics below $T_{pc}$ can be understood in terms of Hadron
Resonance Gas (HRG) models. The main idea behind these models is that the interaction between hadrons in hot matter can be accounted for by
including hadronic resonances, which are treated as stable particles when calculating the partition function. Below the crossover temperature, a HRG model based on resonances listed in tables of
the Particle Data Group (PDG) gives a good
description of the QCD equation of state \cite{HotQCD:2014kol,Borsanyi:2013bia,Bollweg:2022fqq} and fluctuations of conserved charges (BQS) \cite{HotQCD:2012fhj,Bellwied:2015lba,Bollweg:2021vqf,Bollweg:2022rps,Biswas:2024xxh}. We call this HRG model PDG-HRG. Although in some cases it is necessary to include missing hadronic resonances that are not listed in PDG tables but are predicted by the Quark Model (QM) calculations \cite{Bazavov:2014xya,Bollweg:2021vqf,Biswas:2024xxh}. This HRG
model with an extended list of states usually is called QM-HRG. 
The study of the ratios of various generalized charm susceptibilities in the previous lattice QCD studies also
indicated that there are missing charmed meson and baryon states
\cite{Bazavov:2014yba}. Very recently the need for additional charmed hadron states has been discussed 
in the context of the Statistical Hadronization Model extended to include charm effects (SHMc) \cite{Braun-Munzinger:2024ybd}.

Cutoff effects in the generalized charm susceptibilities are large
because of the large charm quark mass. At the same time, the large
charm quark mass leads to poor signal to noise ratio in the calculation of these
quantities. To overcome the problem of poor signal to noise ratios, a
high statistics lattice QCD study of the generalized charm susceptibilities 
has been performed recently using $N_{\tau}=8$ lattices \cite{Bazavov:2023xzm}. This study significantly improved the lattice QCD results on the generalized susceptibilities at low
temperatures and confirmed the assertion that additional charmed
hadrons have to be included in HRG model calculations to arrive at quantitative agreement with lattice QCD results. On the other hand, the behavior of the generalized susceptibilities above the chiral transition can be well understood in a model, where the partial pressure is written as the sum of charm quark partial pressure, charmed meson partial pressure and charmed baryon partial pressure. Within this model charm 
quark appears as a new charm degree of freedom above the crossover
temperature, while charmed hadron-like excitations still give the dominant contribution to the charm pressure up to $T=176$ MeV \cite{Bazavov:2023xzm}. In our previous work, we showed that this model successfully passes numerous validity tests and satisfies various constraints \cite{Bazavov:2023xzm}. We will come back to one of these constraints in Sec. \ref{sec:partial}. The existence of such hadron-like excitations above $T_{pc}$ is also supported by a recent analysis of charmed meson
and baryon correlations functions at non-zero temperature \cite{Aarts:2022krz,Aarts:2023nax}. Furthermore, the generalized charm susceptibilities have been studied in T-matrix microscopic model \cite{Liu:2021rjf} and it was shown that charmed hadron resonances
are needed in order to describe some of these susceptibilities. 

The aim of this paper is to extend our previous study of the generalized
charm susceptibilities by considering finer lattice spacings, corresponding to lattices with temporal
extents, $N_{\tau}=12$ and $16$, and also by considering various combinations 
of generalized charm susceptibilities that will help better understand the nature of charm degrees of freedom across the chiral
crossover temperature. In addition to studying generalized charm
susceptibilities normalised by the quartic charm
fluctuation at $N_\tau=8$, we also show a few $N_\tau=12$ counterparts of these ratios. These results corroborate our previous deduction \cite{Bazavov:2023xzm} that the cutoff effects get largely cancelled in such ratios calculated on relatively coarser lattices. Therefore, it is reasonable to use continuum estimated quartic charm
fluctuation to convert normalised generalized charm susceptibilities calculated on $N_\tau=8$ lattices into continuum estimates of physical observables relevant to open
charm physics. Moreover, we use $N_{\tau}=8, 12$ and $16$ lattices to obtain continuum estimate of the quartic charm fluctuation, $\chi_4^C\equiv \chi_{0004}^{BQSC}$. 

This paper is organized as follows.
In Section II, we discuss various HRG models in the charmed hadron sector and the pressure of an ideal charm quark gas. In Section III, we present the details of the lattice setup used for our lattice QCD calculations, and discuss
cutoff effects in charm susceptibilities and the continuum extrapolation of the quartic charm fluctuation.   
In Section IV, we present our results on various ratios of generalized
susceptibilities and discuss how these support the idea of missing
charmed hadrons.  In Section V, we decompose the charm pressure in terms
of partial pressure of charmed mesons, charmed baryons and charm quarks.
Here we also show how these can be understood in terms of QM-HRG model calculations
below the chiral crossover and discuss the implications of these
findings on the nature of charm degrees of freedom above $T_{pc}$. 
Preliminary results
from these studies have been presented in conference proceedings
\cite{Sharma:2022ztl,Sharma:2024ucs, Sharma:2024edf, Sharma:2025zhe,Sharma:2025gtg}.

\section{Hadron resonance gas models}
\label{sec:hrg}
In a non-interacting HRG model, the QCD pressure can be written as the sum of
the partial pressure of hadrons
carrying open charm degrees of freedom and the partial pressure of hadrons with zero charm quantum number.
Furthermore, the partial pressure of charmed hadrons, $P_C(T,\vec{\mu})$, can be written as the sum of partial pressures
of charmed mesons, $P_M^C$, and charmed baryons, $P_B^C$,
\begin{equation}
{P_C(T,\vec{\mu})=P_M^C(T,\vec{\mu})+P_B^C(T,\vec{\mu})} \text{ .}
\label{eq:P}
\end{equation}
As the masses of charmed mesons and baryons are much larger than the temperatures in the temperature range of interest to us, 
one can use Boltzmann statistics 
and write $P_M^C$ and $P_B^C$ in the following form \cite{Bazavov:2014yba},
\begin{gather}
	\begin{aligned}
		{P_M^C(T,\vec{\mu})}&{=\dfrac{1}{\pi^2}\sum_{i\in \text{C-mesons}}g_i \bigg(\dfrac{m_i}{T}\bigg)^2K_2(m_i/T)} \cdot \\
			&{\text{cosh}(Q_i\hat{\mu}_Q+S_i\hat{\mu}_S+C_i\hat{\mu}_C)} \text{ ,}\\
		{P_B^C(T,\vec{\mu})}&={\dfrac{1}{\pi^2}\sum_{i\in \text{C-baryons}}g_i \bigg(\dfrac{m_i}{T}\bigg)^2K_2(m_i/T)} \cdot \\
			&{\text{cosh}(B_i\hat{\mu}_B+Q_i\hat{\mu}_Q+S_i\hat{\mu}_S+C_i\hat{\mu}_C)} \text{ .}
		\label{eq:McBc}
	\end{aligned}
\end{gather}

Here each term in the sums takes care of 
a charmed hadron with mass $m_i$ and its
anti-particle.
The summation is over all charmed mesons
or baryons, respectively; $g_i$ denotes the degeneracy factors of states with identical mass and quantum numbers;
${K_2(x)}$ is a modified Bessel function of the second kind, which for large arguments is given by ${K_2(x)}\sim\sqrt{\pi/2x}\;e^{-x}\;[1+\mathbb{O}(x^{-1})]$. 

While the sum over charmed baryons
does contain contributions from $|C|=1,\ 2$
and $3$ sectors, the contributions of the $|C|\ne 1$ sectors is strongly suppressed as 
the mass of the only experimentally established (3-star) doubly-charmed baryon ($\Xi_{cc}^{++}$) is roughly about 1.5~GeV larger
than that of the lightest charmed baryon
($\Lambda_c^+$). In the temperature range of
interest to us, {\it i.e.} $T\lsim T_{pc}$,
the ratio $(m_{\Xi_{cc}}-m_{\Lambda_c})/T\sim {\cal O}(10)$ and consequently
the contribution of doubly-charmed baryons to $P_B^C$ is suppressed by a factor $\sim {\rm e}^{-10}$  {\it i.e.} this contribution is less than $0.01\%$ of the charmed baryon pressure.
Similarly, also possibly existing charmed 
baryonic molecules with baryon number $|B|>1$
\cite{Kong:2022rvd,Kong:2023dwz} would be
strongly Boltzmann suppressed. We thus consider
in the following only HRG models for charmed
baryons with $|C|=\pm 1$. Note that our HRG lists contain baryons with $|C|\neq 1$.

We will use in the following two different 
HRG particle lists PDG-HRG2024c and QM-HRG2024c
\cite{datapublication}, which are similar in structure to our earlier lists PDG-HRG and QMHRG2020 containing only non-charmed states \cite{datapublication2020}.  PDG-HRG2024c contains
the recent compilation of all charmed hadrons listed
by the Particle Data Group 
\cite{ParticleDataGroup:2024cfk} and we augmented it in some cases by not observed
isospin partners of experimentally observed
states. For instance, we added the unobserved $D_+^*$ in addition to the observed $D_0^*$. 
To construct the QM-HRG2024c list we used PDG-HRG2024c and added charmed hadron resonances
not observed so far but obtained in quark
model calculations
\cite{Chen:2022asf,Ebert:2011kk,Ebert:2009ua,Roberts:2007ni,Yoshida:2015tia}. More specifically, for not yet observed charmed states we used
the relativistic quark model of Refs. \cite{Ebert:2011kk,Ebert:2009ua}.

In the HRG phase, generalized susceptibilities introduced in Eq.~\eqref{eq:chi} are calculated by making use of the partial pressure expressions introduced above in Eq.~\eqref{eq:McBc}. In particular, for the calculation of generalized charm susceptibilities at vanishing chemical potentials, it suffices to replace the QCD pressure, $P$, with the partial charm pressure, $P_C$. The resulting expression for $\chi^{BQSC}_{klmn}$ takes the following form,
   
\begin{equation}
\chi^{BQSC}_{klmn}=\dfrac{1}{2\pi^2}\sum_{i\in \text{C-hadrons}}g_i \bigg(\dfrac{m_i}{T}\bigg)^2K_2(m_i/T)\; B^kQ^lS^mC^n \text{,}
	\label{eq:chi_HRG}
\end{equation}
where the sum is over all charmed hadrons. 
As discussed above,
the contribution of multiply-charmed  baryons is exponentially small, and effectively only the $|C|=1$ sector contributes
to the pressure. 
This means that $\chi_2^C=\chi_n^C=P_C(T,\vec{\mu})$, for $n$ even, and
$\chi_{11}^{BC}=\chi_{1m}^{BC}=P_B^C$, for $m$ odd. Note that zero subscripts and their corresponding superscripts, {\it e.g.,} $QS$ in $\chi^{BC}_{1m}$, are suppressed, and we will use the same notation in the following.
Above observation motivates our construction for partial 
charm pressures in various quantum number channels. In particular, in the context of non-interacting HRG models, we use simple
cumulants as proxies for the partial pressures of charmed baryon and 
meson sectors,
\begin{eqnarray}
P_B^C(T,\vec{\mu}) &=& \chi_{1n}^{BC}\; ,
 \label{eq:pb}
\\
P_M^C(T,\vec{\mu}) &=& \chi_{n+1}^C - \chi_{1n}^{BC} \; ,\; n\; {\rm odd}\; .
\label{eq:pm}
\end{eqnarray}
For definiteness and in order to be consistent
with the lattice QCD calculations discussed in
the following sections, we will use fourth order 
cumulants, {\it i.e.} we set $n=3$.

\begin{figure}[t]
\includegraphics[width=\linewidth]{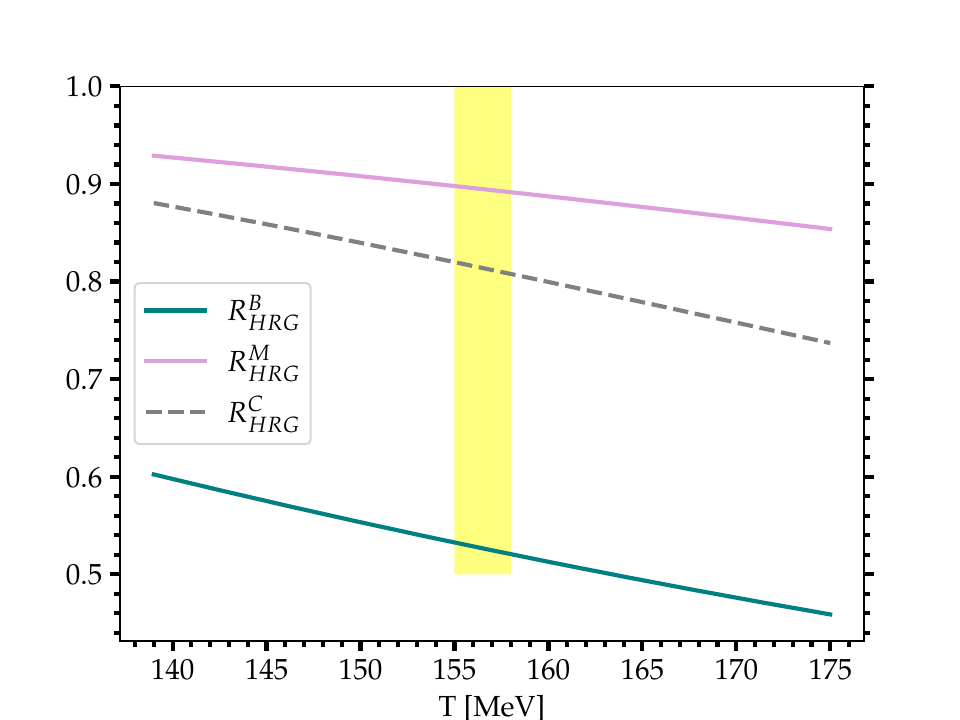}	

\caption{Ratios of PDG-HRG to QM-HRG partial pressure predictions of charmed baryons and mesons, and the total charm pressure as functions of temperature. The yellow band represents $T_{pc}$ with its uncertainty. See eqs. \ref{eq:R} and \ref{eq:RC} for the definitions of $R_{\text{HRG}}^{B/M/C}$.}
\label{fig:barCmesC}
\end{figure}

In Fig. \ref{fig:barCmesC}, we use these proxies of charmed baryon and meson pressures
in the HRG phase, to show the contribution of experimentally observed hadrons to charm pressure in relation to the QM-HRG states. We introduced a temperature dependent ratio, $R_{HRG}^{B/M}$, separately for baryons and mesons, as well as the partial charm pressure
ratio $R_{HRG}^{C}$,
\begin{eqnarray}
    R_{\text{HRG}}^X(T) &=& \frac{P^C_{X}(T)_\text{PDG-HRG}}{P^C_{X}(T)_\text{QM-HRG}} \; ,\; X=B,\ M \label{eq:R}\\
    R_{\text{HRG}}^C(T) &=&
    \frac{P_C(T)_\text{PDG-HRG}}{P_C(T)_\text{QM-HRG}} \; .
    \label{eq:RC}
\end{eqnarray}
The ratio $R_{\text{HRG}}^B(T)$ is around $0.6$ at the lowest temperature, and decreases as the temperature increases. This is because with increasing temperature, the partial pressure contribution of the experimentally undetected heavier resonances, which contribute in QM-HRG calculations, gain in importance at higher temperatures. At $T_{pc}$, $R_{\text{HRG}}^B$ is around $0.52$. 
As we will show in the next sections QM-HRG
calculations are, in fact, in good agreement
with lattice QCD results at and below $T_{pc}$.
This implies that 
at $T_{pc}$, the experimentally known charmed baryons contribute only $52\%$ to the total
charmed baryon pressure of QCD, and a contribution of $48\%$ arises from the not-yet-discovered 
charmed baryonic states. On the other hand, for the charmed mesonic sector, at $T_{pc}$, experimentally unobserved states contribute around $12\%$. As can be seen in Fig.~\ref{fig:barCmesC}, 
the difference between QM-HRG and PDG-HRG partial pressure predictions is much larger in the charmed baryon than in the charmed meson sector.
This is due to the fact that there are many more missing charmed baryonic states than 
missing charmed mesonic states as pointed out 
already in a previous lattice QCD study \cite{Bazavov:2014yba}. In recent years, the 
experimental discovery of various $\Lambda_C$ and $\Omega_C$ resonances, in particular by LHCb and Belle collaborations \cite{Belle:2016tai,LHCb:2017uwr,LHCb:2017jym} have confirmed this. 
As can be seen also in Fig.~\ref{fig:barCmesC}, the ratio for the partial charm pressure, $R^C_\text{HRG}$, is closer to $R^M_\text{HRG}$
reflecting that the partial charm pressure is
dominated by the charmed mesonic contributions. At
$T_{pc}$ we find $P_B^C(T_{pc})= 0.2843 P_M^C(T_{pc})$.

\begin{figure}[t]
	\includegraphics[width=0.45\textwidth]{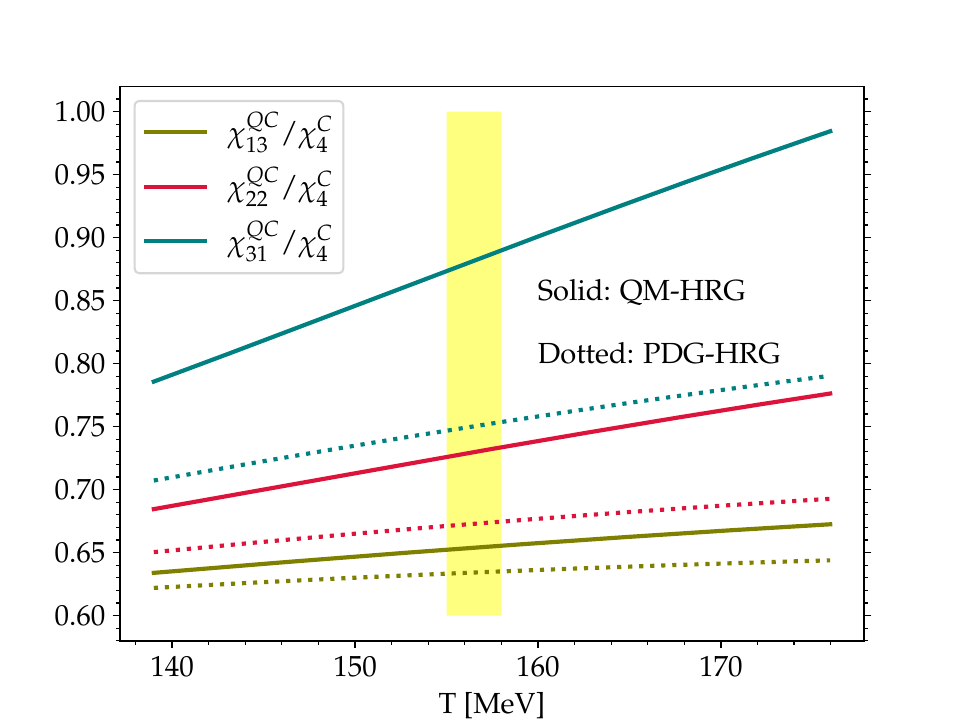}
	\caption{Comparison of QM-HRG and PDG-HRG
	predictions for the ratio of generalized susceptibilities $\chi_{mn}^{QC}/\chi_4^C$. The solid (dotted) lines represent QM-HRG (PDG-HRG) prediction. The yellow band represents $T_{pc}$ with its uncertainty.
	 }
	\label{fig:QC-HRG}
\end{figure}

Recently, the analysis of experimentally measured open charmed hadron yields using the extended Statistical Hadronization Model for charm (SHMc) \cite{Andronic:2018vqh}, showed enhanced yields in the charmed baryon sector \cite{Braun-Munzinger:2024ybd}. This enhancement is relative to the experimentally known charmed hadronic states tabulated in the PDG records, and  therefore corroborates the existence of missing resonances in the open charm sector. 

The fact that differences between QM-HRG and 
PDG-HRG show up predominantly in the charmed baryon sector also is apparent in cumulants
for correlations between electric-charge and charm fluctuations.
As states with $|Q|>1$ show up only in the 
baryon sector, cumulants involving higher order
electric charge fluctuations show larger 
differences between QM-HRG and PDG-HRG calculations. This is evident from  Fig.~\ref{fig:QC-HRG} where we show HRG model calculations for the ratios $\chi_{mn}^{QC}/\chi_4^C$, with $m=1,2,3$ and $n=4-m$. While
the ratio for $m=1$ is dominated by contributions from charmed mesons, the baryon
contribution is enhanced by factor $2^m$ for
larger $m$ leading to larger differences between
QM-HRG and PDG-HRG calculations.

\begin{figure}[t]
\includegraphics[width=\linewidth]{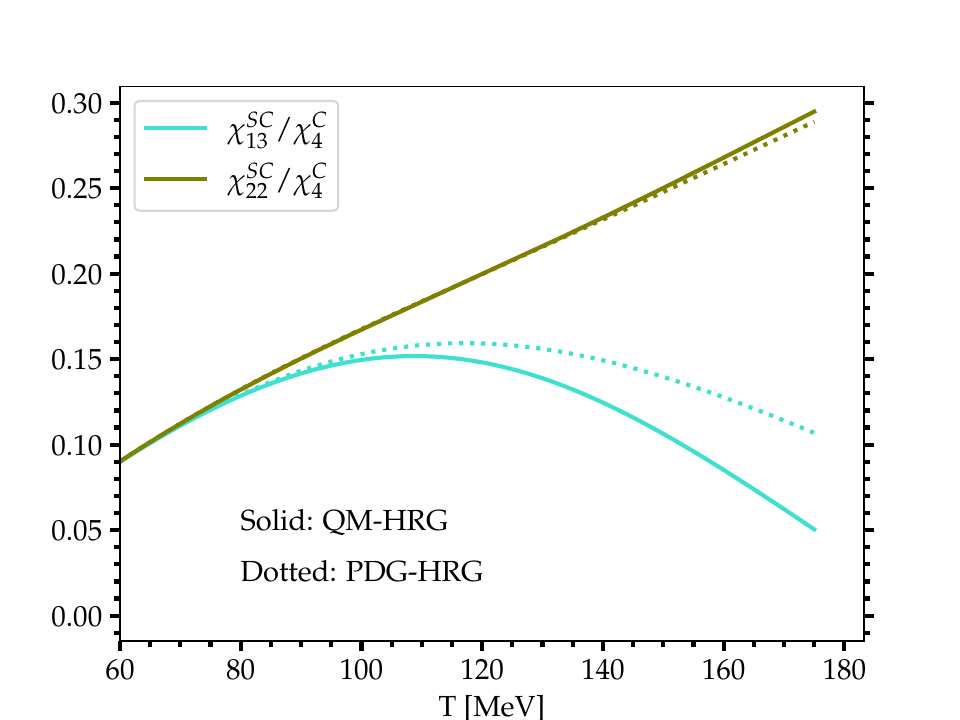}
\caption{Shown are the QM-HRG predictions for the two SC correlations normalised by $\chi^C_4$.}
\label{fig:baryonsSC}
\end{figure}

In Fig. \ref{fig:baryonsSC} we show PDG-HRG and QM-HRG predictions for two strangeness-charm correlations normalized
by $\chi_4^C$. The effect of the  missing states appears to be very small for $\chi_{22}^{SC}/\chi_4^C$,
while it is significant for $\chi_{13}^{SC}/\chi_4^C$. The reason for this is that the missing charmed mesons 
and baryons contribute to both the numerator and denominator of $\chi_{22}^{SC}/\chi_4^C$ and the effect of the missing states
cancels out in the ratio. However, this cancellation is coincidental for the QM-HRG and PDG-HRG lists used in this work. In other words, if in nature the spectrum of strange charmed hadrons,  including the spectrum of $|S|=2$ charmed baryons, is different from  that tabulated in the current QM-HRG list, $\chi_{22}^{SC}/\chi_4^C$ will be sensitive to those states and show deviations from the PDG-HRG prediction. 
On the other hand, charmed mesons and baryons contribute to $\chi_{13}^{SC}$ with opposite signs, therefore, no cancellation between the numerator and denominator takes place
for $\chi_{13}^{SC}/\chi_4^C$. At low temperature values, only charmed mesons contribute to $\chi_{nm}^{SC}$ and therefore,
at these temperatures all strangeness-charm correlations should be the same. We see from  
Fig. \ref{fig:baryonsSC} that this is indeed the case.

We close this section by noting that the 
Boltzmann approximation not only holds for 
charmed hadrons but also is a good approximation
for the pressure of charm quarks even at 
temperatures a few times $T_{pc}$.
The partial pressure, $P_q^C$, of a gas of charm quarks and anti-quarks
with mass $m_q^C$ thus may be written as  

\begin{eqnarray}
	P_q^C(T,\overrightarrow{\mu})&=&\dfrac{6}{\pi^2}\bigg(\dfrac{m_q^C}{T}\bigg)^2K_2(m_q^C/T)\cdot \nonumber \\
	&&\text{cosh}\bigg(\dfrac{2}{3}\hat{\mu}_Q+\dfrac{1}{3}\hat{\mu}_B+\hat{\mu}_C\bigg) \; . 
	\label{eq:pq}
\end{eqnarray}

\section{Lattice setup and lines of constant physics}
 \label{sec:lattice-LCP}

\begin{table}[t]
\centering
\begin{tabular}{|c|c|c|c|c|c|c|}
\hline
~ & ~ & ~ &\multicolumn{2}{c|}{LCP$_{[a]}$}&\multicolumn{2}{c|}{LCP$_{[b]}$}  \\
\hline
$N_\tau$ &$\beta$ & $T$[MeV] & $am_c$ & statistics & $am_c$ & statistics\\
\hline
8  & 6.315 & 145.1 & 1.04112 & 449,689 & 0.892231 & 448,894\\
8  & 6.354 & 151.1 & 0.97025 & 519,953 & 0.857304 & 519,812\\
8  & 6.390 & 156.9 & 0.91534 & 406,878 & 0.816144 & 1,038,000\\
8  & 6.423 & 162.4 & 0.87069 & 661,254 & 0.787450 & 662,607 \\
8  & 6.445 & 166.1 & 0.84320 & 505,573 & 0.765223 & 522,688\\
8  & 6.474 & 171.2 & 0.80920 & 229,013 & 0.742996 & 232,114\\
8  & 6.500 & 175.8 & 0.78059 & --      &0.723946 & 151,478\\
\hline
12 & 6.712 & 145.40 & 0.59316 & 49,589 & 0.574711 & 49,591\\
12 & 6.754 & 151.62 & 0.56328 & 43,367 & 0.549310 &43,368\\
12 & 6.794 & 157.75 & 0.53656 & 50,352 & 0.530258 &50,353\\
12 & 6.825 & 162.65 & 0.51694 & 58,547 & 0.511207 &58,547\\
12 & 6.850 & 166.69 & 0.50176 & 36,801 & 0.498506 &36,803\\
12 & 6.880 & 171.65 & 0.48426 & 36,076 & 0.485805 &36,078\\
12 & 6.910 & 176.73 & 0.46751 & 39,079 & 0.469929 &39,080\\
\hline
16 & 7.010 & 145.95 & 0.41657 &13,546 & 0.419126  & 13,554 \\
16 & 7.054 & 152.19 & 0.39631 &13,391 & 0.409601 &13,390\\
16 & 7.095 & 158.21 & 0.37849 &14,050 & 0.393725 &14,048\\
16 & 7.130 & 163.50 & 0.36405 &6,779 & 0.377849 & 6,807\\
16 & 7.156 & 167.53 & 0.35375 &7,282 & 0.368323 & 7,308 \\
16 & 7.188 & 172.60 & 0.34155 &7,192 & 0.358798 & 7,192  \\
16 & 7.220 & 177.80 & 0.32985 &3,515 & 0.349272 & -  \\
\hline       
\end{tabular} 
\caption{\label{tab:amc} 
The lattice gauge coupling $\beta=10/g_0^2$, the corresponding temperature values and the bare 
charm quark masses for LCP$_{[a]}$ and LCP$_{[b]}$.}

\end{table}
\subsection{Calculation of charmed susceptibilities on lines of constant physics}

For our calculation of 
generalised susceptibilities involving charm quantum numbers we make use of 
gauge field configurations generated 
by the HotQCD collaboration in (2+1)-flavor QCD with a physical strange to light quark mass ratio, ${m_s/m_{l}}=27$ \cite{Bollweg:2021vqf}.
These data sets have been generated using the
(2+1)-flavor QCD  
HISQ (Highly Improved Staggered Quark) 
action and a Symanzik-improved gauge action.
The temperature scale was set using the $f_K$-scale \cite{Bollweg:2021vqf}. 

In our calculations the charm quark sector is treated in the quenched approximation. This approximation is well justified within the temperature range explored in this work, as a comparison between the (2+1)-flavor QCD equation of state (EOS) with quenched charm and the (2+1+1)-flavor QCD EOS including dynamical charm shows that dynamical charm effects become significant only for $T > 300~\text{MeV}$ \cite{Weber:2021hro}.
We use the HISQ action with the so-called epsilon term for charm quarks to remove $\mathcal{O}((am_c)^4)$ tree-level lattice artifacts \cite{Follana:2006rc}.
This setup is known to have small discretization effects for charm quark related observables \cite{Follana:2006rc,MILC:2010pul}.
Tab. \ref{tab:amc} shows the bare gauge coupling
 $\beta=10/g_0^2$ and the number of gauge field configurations utilized for each temperature at three different temporal lattice extents, $N_\tau=8,12, \textrm{ and }16$. The analysed gauge configurations have a fixed aspect ratio, $N_\sigma/N_\tau=4$, where $N_\sigma$ is the spatial lattice extent.
This setup has been used by us previously for the
calculation of partial pressures in the charmed
sector of QCD \cite{Bazavov:2023xzm}.

To calculate the generalized charm susceptibilities we need to fix the input charm quark mass in lattice units, $a m_c$, for each lattice spacing, such
that some physical quantities depending on charm quark mass and expressed in appropriate units of $f_K$ are fixed to their physical values even at finite values of the lattice cut-off. This defines the Lines of Constant Physics (LCPs).
In our previous analysis of charmed susceptibilities we used two different LCPs.
On the one hand we used a LCP obtained by
keeping the spin-averaged charmonium mass, ${(3m_{J/\psi}+m_{\eta_{c\bar{c}}})/4}$, fixed to its physical value \cite{Sharma:2022ztl}. Furthermore, 
we considered also a
LCP defined by the physical (PDG) charm to strange quark mass ratio, $m_c/m_s=11.76$ \cite{ParticleDataGroup:2024cfk}, where bare strange mass values are taken from Ref.~\cite{Bollweg:2021vqf}.  In the following, results based on the above two LCPs will henceforth contain subscripts [a] and [b], respectively.  
The values of the bare charm quark masses corresponding to these two
LCPs are also given in Tab. \ref{tab:amc}.

Comparing the results obtained using different LCPs for the charm quark mass helps to understand the cutoff effects in the
calculations of the generalized charm susceptibilities. E.g. it is expected that
cutoff effects are large in calculations
of thermodynamic observables dominated by charm quark contributions as
the quark mass is an order of magnitude
larger than the temperatures of interest.
Thus even small changes in the quark
masses may lead to large changes in the
relevant Boltzmann weights controlling
the contribution of charmed hadrons in 
thermodynamic calculations of, e.g.
absolute values of generalized susceptibilities.
This will, in fact, become clear from our discussion in subsequent sections. For this reason previous analyses of charmed susceptibilities focused on the calculation of ratios of susceptibilities, where such 
cut-off effects are canceled to a large extent.

As we want to arrive here at quantitative results for absolute values of continuum extrapolated charmed susceptibilities
which at low temperatures are dominated
by contributions from open charmed mesons,
we constructed a third LCP, which aims at
reducing cut-off effects in these charmed hadron degrees of freedom. We thus 
will discuss in the next subsection the construction of a LCP obtained by keeping
the lowest $D$-meson mass constant.
 
 \begin{figure}[t]
	
\includegraphics[width=\linewidth]{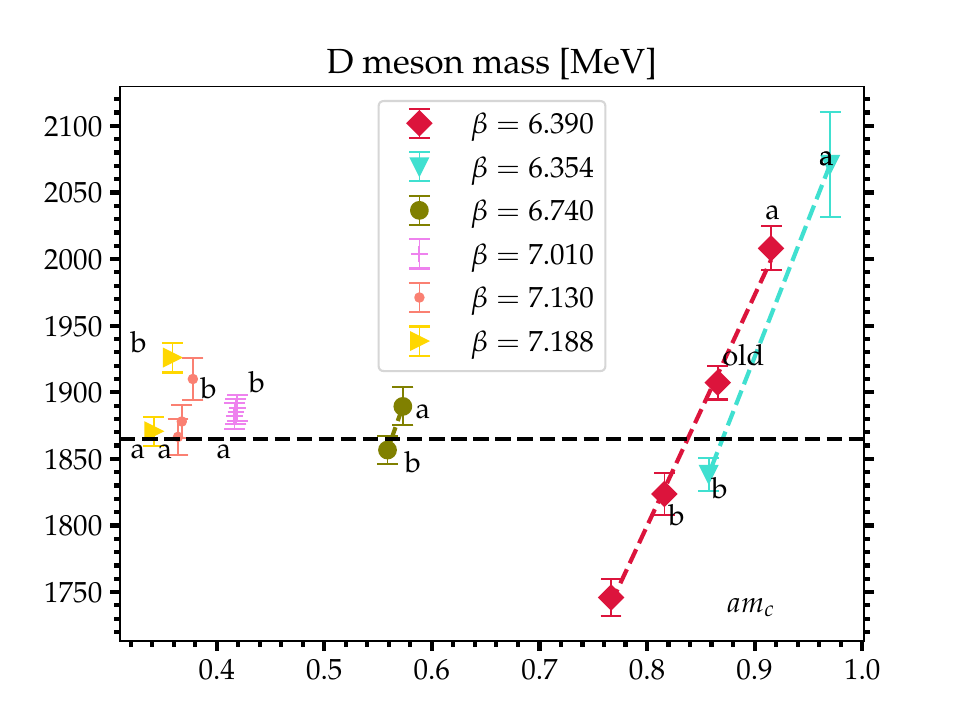}
\caption{The mass of the $D$-meson calculated at several values of the gauge coupling $\beta$ and various values of the bare charm quark mass $am_c$ on LCP$_{[a]}$, LCP$_{[b]}$ and an older LCP introduced in Ref.~\cite{Bazavov:2014yba}. The dashed black line represents the physical $D$-meson mass $= 1864.84$ MeV.}
\label{fig:Dmass}
\end{figure}
 
 \begin{figure}[t]	
	\includegraphics[width=\linewidth]{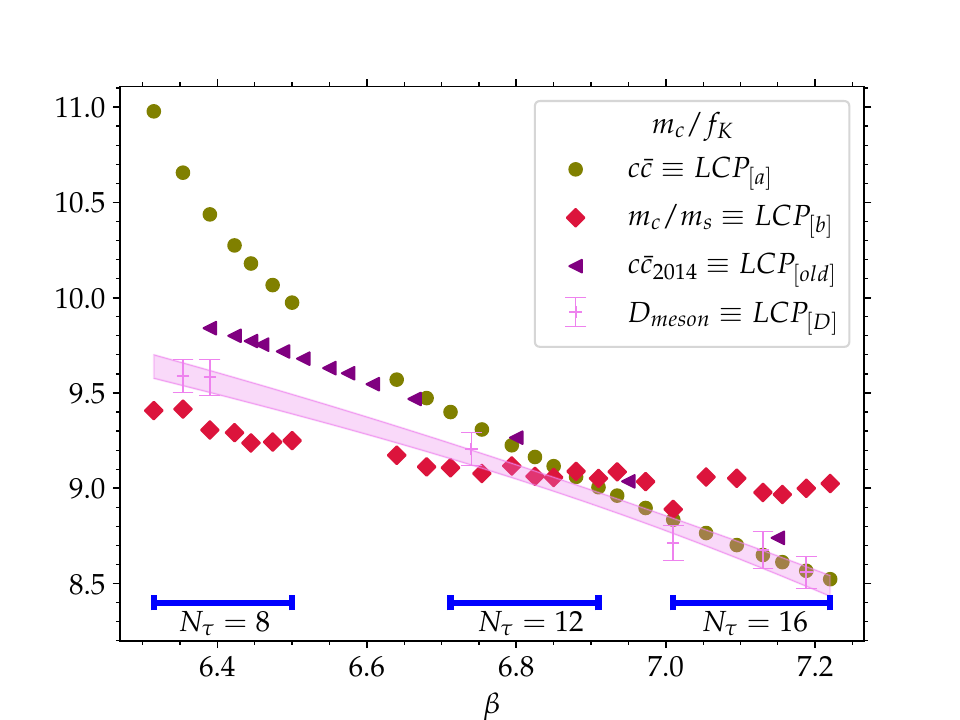}
	\caption{Shown are the bare charm quark mass values normalised by $af_{K}$ as a function of the inverse gauge coupling, $\beta$, tuned using four different criteria to define lines of constant physics: LCP$_{[a]}$, LCP$_{[b]}$ (see text), the LCP prescription used in Ref.~\cite{Bazavov:2014yba}, and LCP$_{[D]}$ constructed from Fig. \ref{fig:Dmass} (see text). The band shows bootstrap error of LCP$_{[D]}$ values. The blue lines explicitly show the ranges of gauge couplings $\beta$ relevant for three different temporal lattice extents used in this work.}
	\label{fig:para_comp}
	\end{figure}

\subsection{Line of constant physics from $D$-meson mass calculations}

In order to minimize cut-off effects that can
arise in thermodynamics of charmed hadrons
from taste breaking lattice artifacts at 
low temperature -- the hadronic phase of QCD --
we construct a LCP obtained by demanding that the mass of the lightest charmed hadron, {\it i.e.,} $D$-meson, attains its physical value already 
at non-zero values of the lattice spacing.
We calculated the lightest $D$-meson mass
at a few $\beta$ values, with at least one $\beta$ value belonging to the parameter range 
of relevance for thermodynamic calculations at each of the three $N_\tau$ values used in this work, $N_\tau=8,12$ and $16$. 
We performed calculations with bare charm quark mass values $am_c$ used in previous 
calculations for the construction of LCP$_{[a]}$
and LCP$_{[b]}$ in the range $\beta\le 6.5$ \cite{Bazavov:2023xzm} and
constructed these LCPs also for larger $\beta$-values 
relevant for calculations on $N_\tau=12$ and $16$. The resulting bare charm quark masses are given in Tab~\ref{tab:amc}. 
We also used the charm quark mass values 
used for the LCP in Ref.~\cite{Bazavov:2014yba}.
We selected
six $\beta$-values in the range $\beta \in [6.39:7.188]$ at which we calculated the $D$-meson mass. For this purpose we used (2+1)-flavor zero temperature lattices of size $32^4$ for $\beta=6.354 \textrm{ and } 6.390$, $48^4$ lattices for $\beta=6.740$, and $64^4$ lattices for $\beta \ge 7.010$. 
At $\beta=6.39$, $\beta=7.010$ and $\beta=7.13$
we also performed calculations with an additional choice for $am_c$. Results for the $D$-meson mass
obtained at these six different $\beta$-values and 
the various charm quark masses are shown in
Fig.~\ref{fig:Dmass}.
The lowest two $\beta$ values shown in this figure are relevant for $N_\tau=8$, whereas the highest three are relevant for $N_\tau=16$, and $\beta=6.740$ is used in $N_\tau=12$ calculation. 

For each $\beta$ value shown in  Fig.~\ref{fig:Dmass}, we fitted the available $D$-meson masses as linear function of $am_c$, and obtained $am_c$ values corresponding to the physical $D$-meson mass. 
The resulting six $am_c/af_K$ values, corresponding to physical $D$-meson mass, are shown for a comparison with other LCPs in Fig.~\ref{fig:para_comp}, and are labelled as LCP$_{[D]}$. The errors on these six $am_c$ values represent the standard deviation of three different linear fits: 1) all $D$-meson masses are moved up one sigma, 2) all $D$-meson masses are moved down one sigma, 3) fitting the original (non-displaced) $D$-meson masses. Finally, we fitted these six $am_c$ values using a renormalisation group inspired ansatz,

\begin{figure*}[t]
	\includegraphics[width=0.34\linewidth]{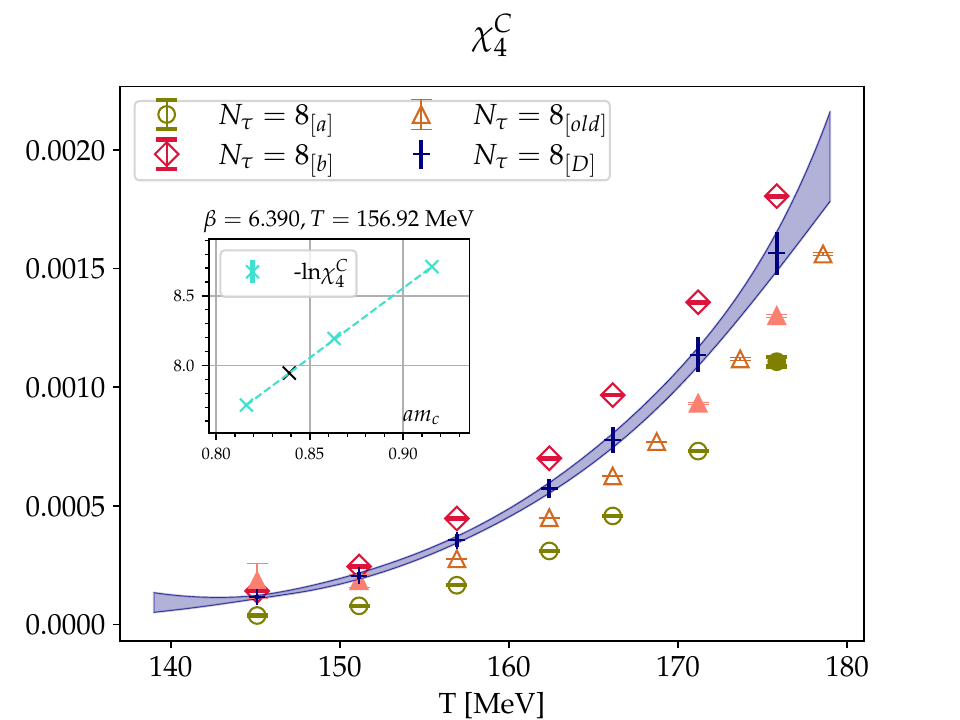}\hspace{-0.3cm}
	\includegraphics[width=0.34\linewidth]{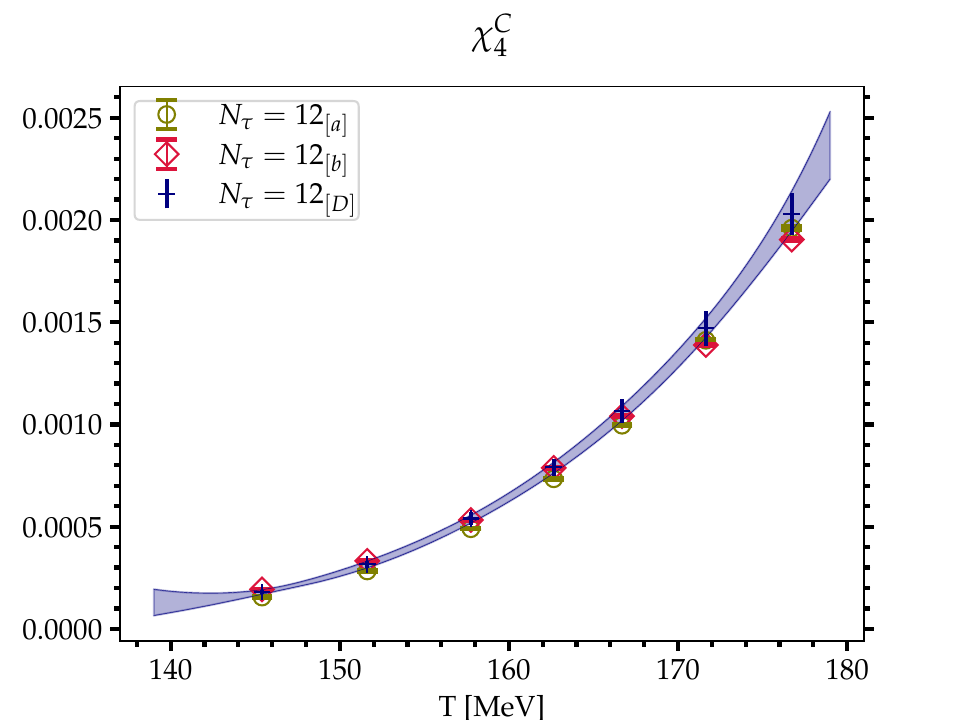}\hspace{-0.3cm}
	\includegraphics[width=0.33\linewidth]{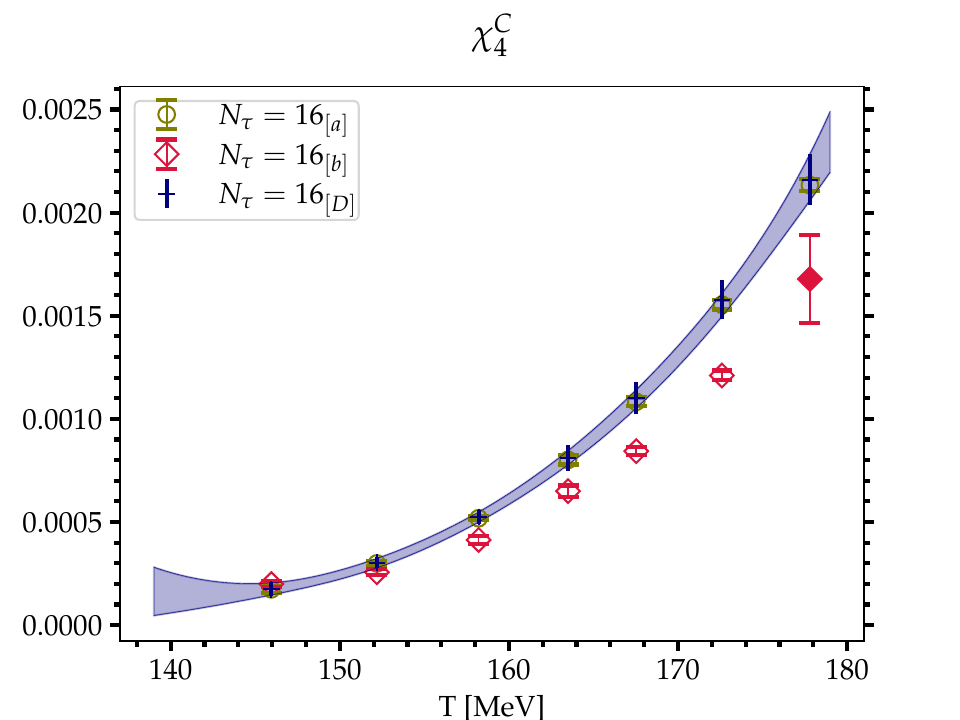}
	\caption{Shown are the charm pressures constructed for LCP$_{[D]}$ as functions of temperature for $N_\tau=8,12, 16$ (see text). The solid points and the bands shown are results of $[2, 2]$ Pad\'e interpolations. The inset in the ({\it left}) figure shows linear interpolation of $-\textrm{ln } \chi^C_4$ in $am_c$ at $\beta=6.390$ and the black cross represents $-\textrm{ln } \chi^C_4$ on LCP$_{[D]}$.  }
	\label{fig:LCP_D}
	\end{figure*}

\begin{equation}
 am_c(\beta) = af_K(\beta) \dfrac{M^{\textrm{RGI}}_c}{f_K} \bigg(\dfrac{20b_0}{\beta}\bigg)^{\frac{4}{9}} \Bigg[\dfrac{1+\frac{10}{\beta} m_{1c} f^2(\beta)}{1+\frac{10}{\beta} dm_{1c} f^2(\beta)}\Bigg] \text{ ,}
	\label{eq:amcD}
\end{equation}
where $af_K(\beta)$ denotes the parametrization of the kaon decay constant times lattice
spacing ($af_K$) given in \cite{Bollweg:2021vqf} and
$f(\beta)$ is the 2-loop QCD $\beta$-function with $b_0=9/(16\pi^2)$ being its 1-loop perturbative expansion coefficient in 3-flavor QCD. We use $M^{\textrm{RGI}}_c = 1.528$ GeV
and $f_K=(155.7/\sqrt{2})$~MeV taken from
Eqs.~58 and 80 of Ref.~\cite{FlavourLatticeAveragingGroupFLAG:2024oxs}, respectively. The fit resulted in 
\begin{eqnarray}
m_{1c} &=& 140585.579\pm 16567.799 \; , \nonumber \\
dm_{1c} &=& 92506.328\pm 11434.443 \; ,
\label{eq:d1}
\end{eqnarray}
with a $\chi^2/dof=0.414$. In Fig.~\ref{fig:para_comp}, the band represents the bootstrap error of fits performed on 100 Gaussian distributed fake samples using Eq.~\ref{eq:amcD}.  

In addition to the LCP$_{[D]}$, shown in Fig.~\ref{fig:para_comp} as band, 
we show there also the values for
$m_c/f_K$ for
LCP$_{[a]}$ and LCP$_{[b]}$ and
the `old' LCP taken from Ref.~\cite{Bazavov:2014yba}, which is also based on the spin-averaged charmonium mass calculations. 
 We note that the LCPs based
on charmonium mass calculations and $D$-meson mass
calculations start agreeing with each 
other in the continuum limit, {\it i.e.}
for large values of $\beta$.  For $\beta\ge 6.74$ differences between charm quark masses on LCP$_{[a]}$ and LCP$_{[D]}$
are less than 2\% leading also to deviations of the $D$-meson mass on LCP$_{[a]}$ from the physical value that
are smaller than 2\%. The difference between LCP$_{[a]}$ and LCP$_{[D]}$ in the $\beta$ range relevant for $N_\tau=8$ lattices is due to the taste symmetry breaking associated with the staggered fermion formulation. Therefore, if the charm quark mass is fixed through charmonium mass at a given lattice spacing 
 then the D-meson mass will be larger than in the continuum limit and vice versa, see Ref.~\cite{HotQCD:2014kol}. 
 There are also discretization errors due to relatively large values of $a m_c$, however, these are under control, due to the use of HISQ action, 
 for which these effects start at order $(a m_c)^4$. The masses of $\chi_c$ and $D_s$ mesons, that are
 not affected by taste symmetry breaking a lot, turned out to agree with the experimental values
 within errors (see {\it e.g.} Ref.~\cite{Bazavov:2014cta}), indicating that the discretization effects proportional to powers of $a m_c$ are small.

As can be seen in Fig.~\ref{fig:para_comp}, 
LCP$_{[b]}$ also approaches LCP$_{[D]}$
with increasing $\beta$.
However, for $\beta\ge 6.9$ LCP$_{[b]}$ results in significantly larger values of
the charm quark mass.
This is due to the fact the LCP used in (2+1)-flavor QCD calculations  with the 
HISQ action \cite{Bollweg:2021vqf,HotQCD:2014kol}
had been obtained by keeping the strange
$\eta_{s\bar{s}}$ meson mass fixed to a value, which is (2-3)\% larger than its physical value. This is discussed in detail in Ref. \cite{HotQCD:2014kol}
and resulted in a 
correspondingly larger strange quark 
mass, which thus is not suitable for
fixing the charm quark mass at large
values of $\beta$ by demanding a 
physical value for the ratio $m_c/m_s$
The resulting charm quark mass is too
large in this region resulting in too
large values for the $D$-meson mass.
This is reflected in the ordering of
the $D$-meson masses calculated with
$am_c$ values corresponding to LCP$_{[a]}$ and  LCP$_{[b]}$. As can be
seen in Fig.~\ref{fig:Dmass}, for 
$\beta\gsim 7.0$ the $D$-meson masses using as input the $am_c$ values of  LCP$_{[b]}$ start being larger than
those based on input from LCP$_{[a]}$.

In the $\beta$ range relevant for $N_{\tau}>12$, LCP$_{[a]}$ is effectively the LCP$_{[D]}$. Since the generalized susceptibilities receive the most dominant contribution from the lightest charmed hadron, we use LCP$_{[D]}$ results to take the continuum limit.

\section{Lattice QCD calculations of generalized charm susceptibilities}

\label{sec:continuum}

\subsection{Quartic charm fluctuations on
\boldmath LCP$_{[D]}$}

In the previous section we have obtained four different LCPs. 
On three of them various charm susceptibilities have been calculated directly. 
On LCP$_{[D]}$ this is not the case as 
its construction is based on the calculation of charmed meson masses at only 6 values of the gauge couplings 
and subsequent fits to the resulting $am_c(\beta)$ values given in Eqs.~\ref{eq:amcD} and \ref{eq:d1}. 
We now could go ahead and perform a direct calculation of 
charm cumulants at these bare charm quark mass values. However, in order
to gain further insight into the sensitivity of quark mass and cut-off effects on the charmed cumulants we discuss here
their determination through interpolation
of data obtained on the other three LCPs.

At each temperature value given in Tab.~\ref{tab:amc}, we made use of the available bare charm masses  
to interpolate $\text{ln}(\chi^C_4)$ as a linear function of $am_c$. This linear interpolation enabled us to eventually obtain $\chi^C_4$ as a function of temperature on LCP$_{[D]}$ using $am_c$ parametrization defined in Eq.~\ref{eq:amcD}.

For $N_\tau=8$, we have at each temperature three values of $am_c(\beta)$
corresponding to LCP$_{[a]}$, LCP$_{[b]}$ and LCP$_{[old]}$, respectively. The quartic charm fluctuation cumulant, $\chi_4^C$, for these three values of $am_c$ are shown in
Fig.~\ref{fig:LCP_D}~({\it left}).
The inset in this figure shows $-\ln(\chi_4^C)$ for one value of $T$.
As can be seen a straight line interpolation is very well justified.
As expected
$\chi_4^C$ receives its dominant contribution from the lightest charmed hadron or charm quark, respectively. 
In both cases these thermal excitations
are exponentially suppressed and the
exponent is given in terms of a thermal 
mass that  is linearly dependent on $am_c$, as shown in Fig.~\ref{fig:Dmass}. 
We thus use linear interpolations of 
results for $-\ln(\chi_4^C)$ to 
determine $\chi^C_4$ values on LCP$_{[D]}$. Results for $\chi^C_4$ are shown in
Fig.~\ref{fig:LCP_D}~({\it left})
as blue plus markers.
The errors shown for $\chi^C_4$ on LCP$_{[D]}$, incorporate not only the bootstrap error arising from the linear fitting process but also the propagated error due to the uncertainties in the $am_c$ values of LCP$_{[D]}$. The propagated error at a given temperature is calculated as 
$\Delta am_{c_{[D]}}(\chi^{C}_{4_{[b]}}-\chi^{C}_{4_{[a]}})/(am_{c_{[a]}}-am_{c_{[b]}})$, where  $\Delta am_{c{[D]}}$ is obtained from the band shown in Fig.~\ref{fig:para_comp}. In Fig~\ref{fig:LCP_D}~({\it left}), the band for LCP$_{[D]}$ is constructed by bootstrapping over 100 fake Gaussian samples, and for each sample a $[2, 2]$ Pad\'e was used as fit function. We also used a $[2, 2]$ Pad\'e to interpolate in temperature and calculate the bootstrap error of $\chi^C_4$ at temperatures where data was missing for any of the three LCPs. 
We repeated the same procedure for $N_\tau=12 \text{ and } 16$, where data for only two LCPs is available, as is shown in Fig~\ref{fig:LCP_D}~({\it middle}) and ({\it right}). 
However, as can be seen in Fig.~\ref{fig:para_comp}, at the
the lowest temperature of $N_\tau=16$, both LCP$_{[a]}$ and LCP$_{[b]}$ 
the charmed quark masses determined for
these two LCPs are almost identical,
which made a meaningful interpolation
difficult. At this lowest temperature we

used LCP$_{[a]}$ result to estimate $\chi^C_4$ corresponding to LCP$_{[D]}$ in Fig~\ref{fig:LCP_D} ({\it right}).

\begin{figure}[t]
	\includegraphics[width=\linewidth]{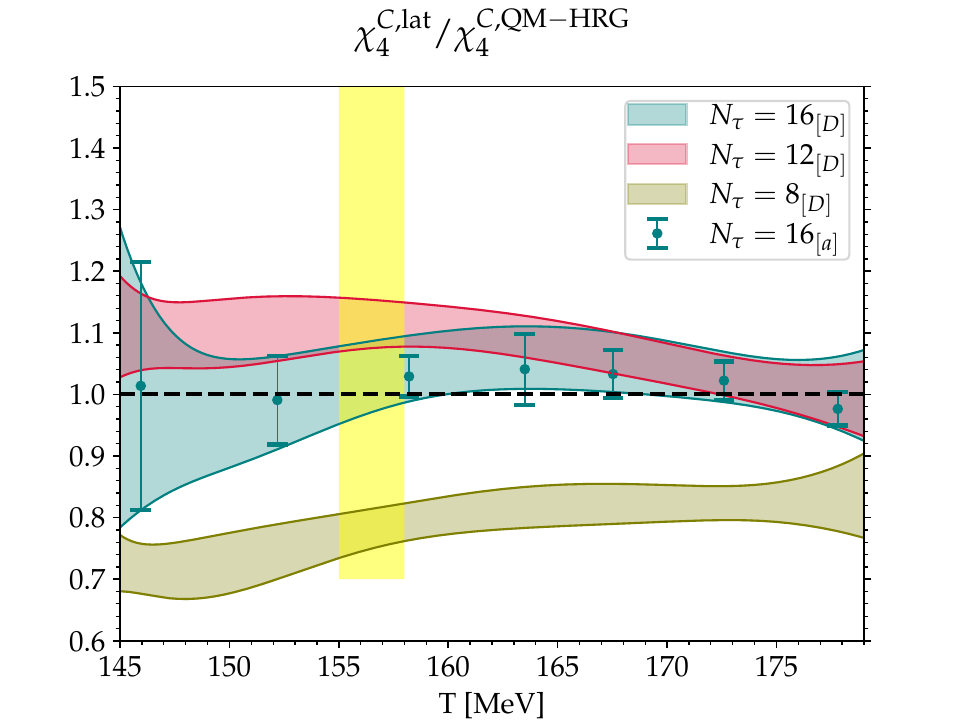}
	\caption{Shown are $\chi^C_4$ values for LCP$_{[D]}$ normalised by the respective QM-HRG predictions as functions of temperature for three different temporal lattice extents. $\chi^C_4$ data points for $N_\tau=16$ on  LCP$_{[a]}$ are also shown. The dashed black line represents unity. The yellow band represents $T_{pc}$ with its uncertainty. For the description of $\chi^C_4$ bands, see text.}
	\label{fig:cont}
	\end{figure}

In Fig.~\ref{fig:cont}, we present $\chi^C_4$ calculated on LCP$_{[D]}$ for $N_\tau=8,12\text{ and } 16$. These values, denoted as $\chi^{C, \textrm{lat}}_4$, are normalized by the QM-HRG value, $\chi^{C, \textrm{QM-HRG}}_4$. As can be seen, LCP$_{[D]}$ bands for $N_\tau=12 \text{ and } {16}$ overlap. Therefore, we use $N_\tau=16_{[D]}$ as our continuum estimate. 
We also see from the figure that our continuum estimate for $\chi_4^C$ agrees within errors with the QM-HRG result for this cumulant in the entire temperature range considered in this study.

 The fact that HRG model calculations seem to describe the lattice
 results above $T_{pc}$ may appear somewhat unexpected. We think this is accidental.
 Both quark gas and hadron gas show qualitatively similar exponential increase
 of $\chi_4^C$ with increasing temperature since charm quark mass is not too
 different from the hadron masses. Given that we only have a crossover not
 a phase transition, $\chi_4^C$ may not be very sensitive to the change in the
 degrees of freedom. In the next section, we consider different
 ratios of generalized susceptibilities, some of which are very sensitive to the change in charm
 degrees of freedom across $T_{pc}$. We will show there that sensitivity to the change in the charm
 degrees of freedom depends on the value of these ratios in the high temperature limit.

\begin{figure}[t]
\includegraphics[width=\linewidth]{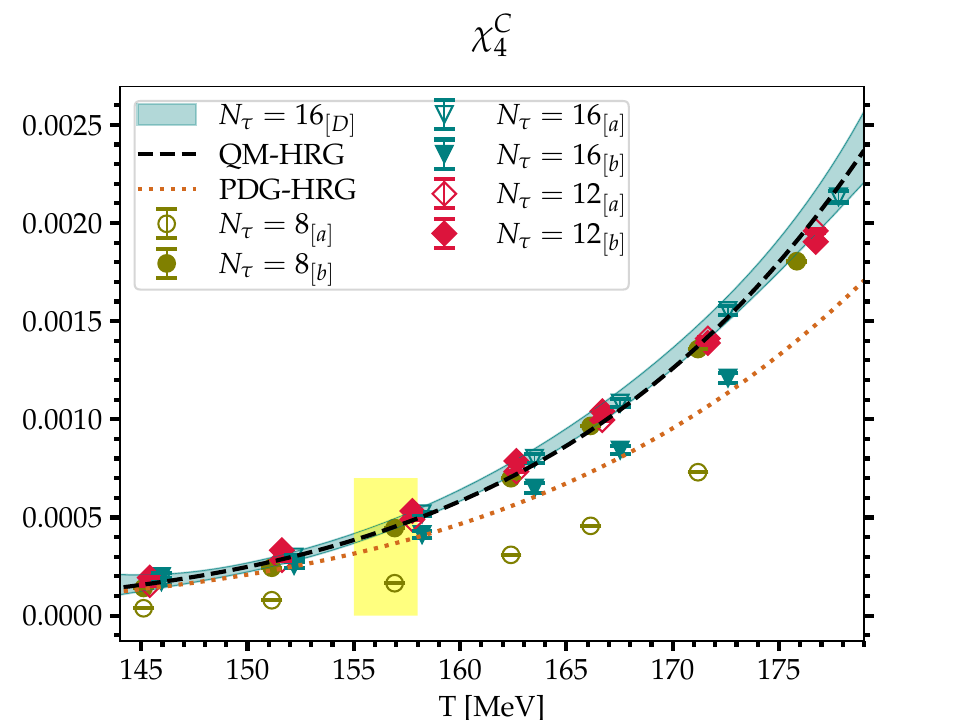}
\caption{Lattice QCD results for 
$\chi_4^C$ for different values of the temporal lattice extents calculated on the two different LCPs, {\it i.e.} LCP$_{[a]}$ and LCP$_{[b]}$ respectively. Also shown are the QM-HRG and the PDG-HRG predictions. The teal band shows $\chi_4^C$ on LCP$_{[D]}$ for $N_\tau=16$, which also represents our continuum estimate of $\chi_4^C$. The yellow band represents $T_{pc}$ with its uncertainty.}
\label{fig:c4}
\end{figure}

\subsection{Lattice cutoff effects in generalized charm susceptibilities}

In Fig.~\ref{fig:c4} we show results
for $\chi_4^C$ obtained by us 
at all bare charm quark mass values corresponding to LCP$_{[a]}$,
LCP$_{[b]}$ and the continuum estimate
for  $\chi_4^C$ on LCP$_{[D]}$.
As can be seen results obtained
with quark masses $am_c(\beta)$, corresponding to
LCP$_{[b]}$ and LCP$_{[D]}$ in the 
parameter range relevant on $N_\tau=8$
lattices agree well, while results
on LCP$_{[a]}$ differ quite significantly. This discrepancy just 
reflects the large difference in 
quark mass values corresponding to
LCP$_{[a]}$ and LCP$_{[D]}$, respectively. This clearly reflects
the importance of cut-off effects arising
from different physical criteria used 
to define LCPs on such coarse lattice,
{\it i.e.} fixing the physical value
of the charmonium mass (LCP$_{[a]}$)
versus fixing the lightest $D$-meson mass
(LCP$_{[D]}$). As can be seen, these 
differences are gone for values 
of $am_c(\beta)$ in the 
parameter range relevant on $N_\tau=12$
lattices. All resulting $\chi_4^C$
values agree within errors despite the 
fact that the input values for $am_c(\beta)$ at a given temperature value differ slightly, {\it i.e.} by about 3\%.

Larger difference between results obtained 
at quark mass values corresponding 
to LCP$_{[b]}$ and the two other LCPs 
show up again on $N_\tau=16$ lattices.
These however, can be traced back to 
discrepancies arising from the strange 
quark mass tuning for $\beta\ge 6.9$
as discussed in the previous section.

From the comparison of results obtained
with different values of the bare charm quark 
masses shown in Fig.~\ref{fig:c4} we 
conclude that input quark masses
corresponding to 
LCP$_{[b]}$ yield
results for $\chi_4^C$ that agree
with results obtained on LCP$_{[D]}$
in the parameter range relevant
for calculations on $N_\tau=8$ and 12 lattices. In the discussion of other 
quartic charm cumulants we thus will use results for these cumulants directly
obtained with bare quark masses 
corresponding to LCP$_{[b]}$, as given
in Tab.~\ref{tab:amc}, rather than performing 
interpolations to obtain results 
on LCP$_{[D]}$ as we have done for $\chi_4^C$.

\section{Ratio of generalized susceptibilities across the chiral crossover}

In this section, we discuss ratios of off-diagonal charm susceptibilities
to $\chi_4^C$ as functions of temperature across the chiral crossover using results from calculations on $N_{\tau}=8$ and $N_{\tau}=12$ lattices.
For $T<177$ MeV we will use $am_c(\beta)$ parameters corresponding to LCP$_{[b]}$, while for larger temperatures,
we will use $N_{\tau}=8$
results for these quantities calculated on  LCP$_{[old]}$ \cite{Bazavov:2014yba}. 

Analyzing the ratios of generalized susceptibilities offers the advantage of significantly reducing sensitivity to the choice of LCP \cite{Bazavov:2023xzm} and
as we will see later, discretization errors become insignificant.
The latter is important because for many generalized susceptibilities
the lattice QCD results for $N_{\tau}=12$ and $N_{\tau}=16$ are too noisy.

\begin{figure}
    \centering
    \includegraphics[width=0.45\textwidth]{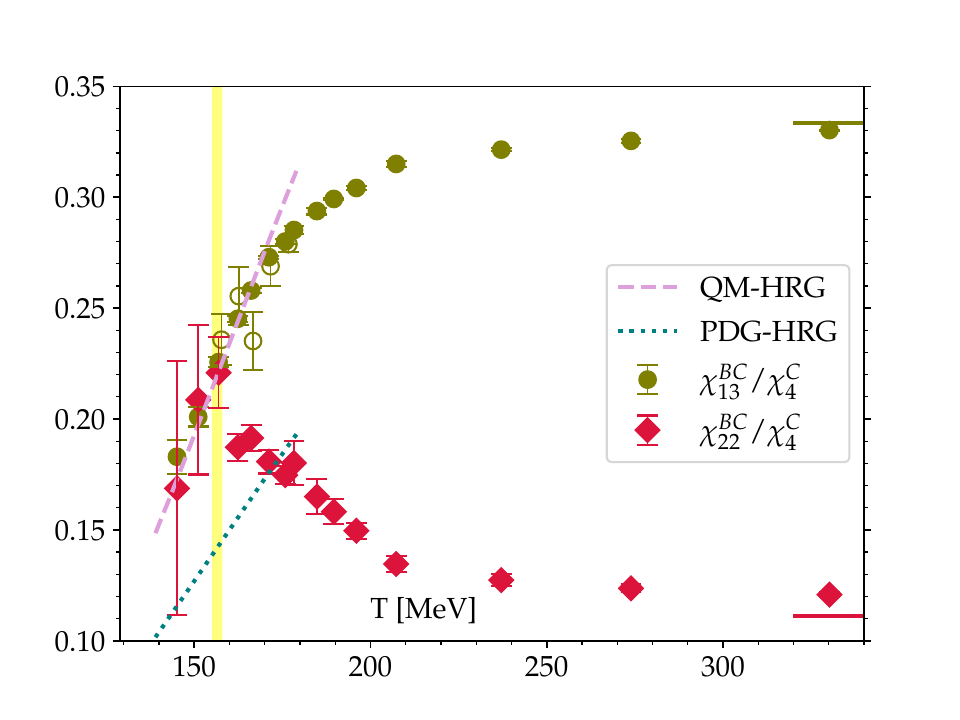}
    \caption{The ratios $\chi_{mn}^{BC}/\chi_4^C$ as function of the temperature
    compared to QM-HRG and PDG-HRG as well as to the ideal quark gas predictions
    at high temperatures shown as horizontal lines.  Solid markers represent $N_{\tau}=8_{[b]}$ results, whereas unfilled markers of the same color represent the respective $N_{\tau}=12_{[b]}$ results. Solid markers for $T>176$~MeV represent $N_{\tau}=8_{[old]}$ results taken from Ref. \cite{Bazavov:2014yba}. Solid lines represent non-interacting charm quark gas limits. The vertical yellow band indicates the chiral crossover temperature with the corresponding uncertainty. }
    \label{fig:BC}
\end{figure}

In our previous study we only considered
$\chi_{13}^{BC}/\chi_4^C$  \cite{Bazavov:2023xzm}. We will extent 
this study here by investigating further
cross-correlations which allows us to
become more sensitive to different
quantum number channels correlated with
charm degrees of freedom.
We start our discussion with the ratios of
baryon number charm correlations to the quartic charm fluctuation, 
$\chi_{mn}^{BC}/\chi_4^C$. The corresponding
lattice QCD results are shown in Fig.~\ref{fig:BC}.
For $T<T_{pc}$, the ratios $\chi_{13}^{BC}/\chi_4^C$ and $\chi_{22}^{BC}/\chi_4^C$
agree with each other as expected based on the Eq.~\eqref{eq:pb}. Furthermore, these ratios agree with HRG model calculations based on the QM-HRG particle list but clearly
disagree with calculations based on the PDG-HRG list. The ratio $\chi_{13}^{BC}/\chi_4^C$ agrees with QM-HRG
prediction even for $T>T_{pc}$. This agreement is somewhat accidental and is
related to the fact that the high temperature ideal quark gas limit of this ratio is 1/3. The ratio $\chi_{22}^{BC}/\chi_4^C$ which is expected to approach
1/9 at high temperature (ideal quark gas value) shows clear disagreement with the QM-HRG above $T_{pc}$. For $\chi_{13}^{BC}/\chi_4^C$ we also show $N_{\tau}=12$ results for a smaller temperature range, $[T_{pc}:177~{\rm  MeV}]$, calculated on LCP$_{[b]}$. These results agree well with the $N_{\tau}=8_{[b]}$ results, showing
that the cutoff effects cancel out in the ratios of generalized susceptibilities.

\begin{figure}[ht]
\includegraphics[width=0.48\textwidth]{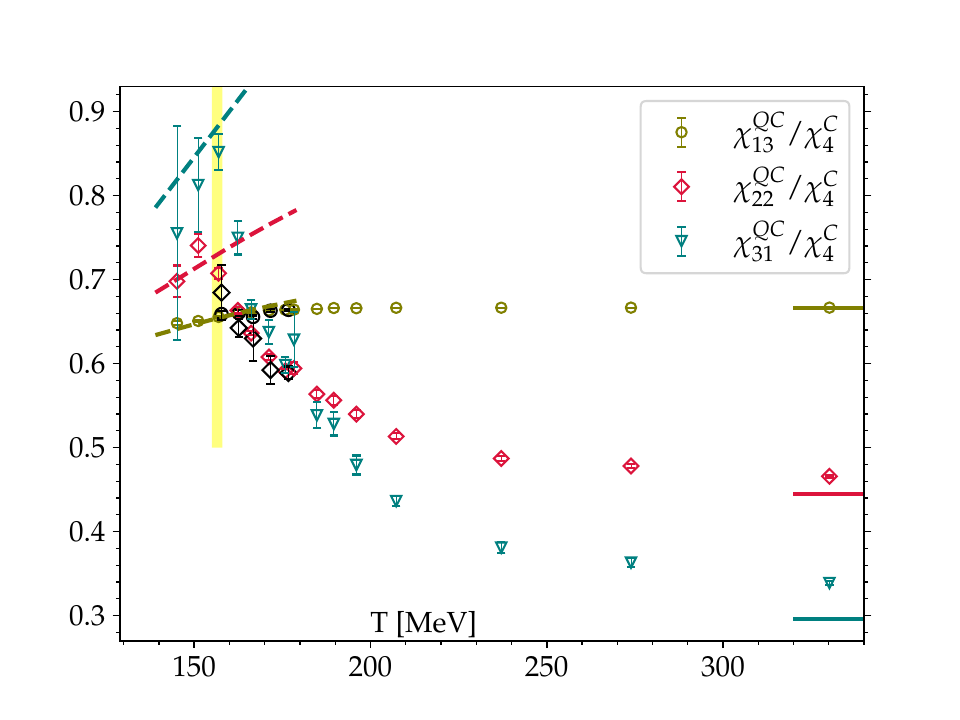}
	\caption{Lattice QCD results for three
	different fourth order cumulants of net electric charge and net charm correlations. The cumulants
	have been normalized using the proxy for the total
	charm pressure, $\chi_4^C$. The unfilled colored markers represent $N_{\tau}=8_{[b]}$ results, whereas the black markers of same style represent the respective $N_{\tau}=12_{[b]}$ results. The unfilled colored markers for $T>176$ MeV represent $N_{\tau}=8_{[old]}$ results taken from Ref. \cite{Bazavov:2014yba}.
	The corresponding QM-HRG predictions are shown with the dashed lines of the same colors, whereas the horizontal solid lines on the left hand side represent the respective ideal charm quark gas limits. 
	The yellow bands represent $T_{pc}$ with its uncertainty. 
	 }
	\label{fig:QC}
\end{figure}

We also show ratios of electric charge charm correlations to the quartic charm fluctuation, $\chi_{mn}^{QC}/\chi_4^C$, in Fig.~\ref{fig:QC}. As can be seen from Tab.~\ref{tab:amc}, these quantities calculated on the high statistics data sets, enable disentangling contributions to partial charm pressure from different electrically charged charm sub-sectors {\it i.e.,} $|Q|=0,1,2$ and $1/3$ in the hadronic and QGP phases, respectively. The fourth-order $QC$ correlations normalised by the total charm pressure show agreement with the QM-HRG model calculations below $T_{pc}$ for $N_\tau=8$ lattices. This indicates the existence of experimentally unobserved charmed hadrons carrying $|Q|=0,1 \text{ and } 2$. We also show $N_\tau=12$ results for ratios $\chi^{QC}_{22}/\chi^C_4$ and $\chi^{QC}_{13}/\chi^C_4$, and despite being relatively noisy and based on lower statistics, these agree with their $N_\tau=8$ counterparts calculated on the same LCP. Therefore, similar to the $BC$ sector, the cutoff effects also cancel for ratios of generalized susceptibilities in the $QC$ sector. We also see that $\chi_{13}^{QC}$ follows the QM-HRG model even above crossover temperature -- again somewhat
accidental. The ideal quark gas limit of this quantity is not very different from the corresponding QM-HRG
value close to $T_{pc}$. The other correlations normalised by $\chi_4^C$, {\it i.e.} $\chi_{31}^{QC}/\chi_4^C$ and $\chi_{22}^{QC}/\chi_4^C$, have much smaller ideal
quark gas limits, and therefore show significant deviations from QM-HRG just above $T_{pc}$.

\begin{figure}
\includegraphics[width=\linewidth]{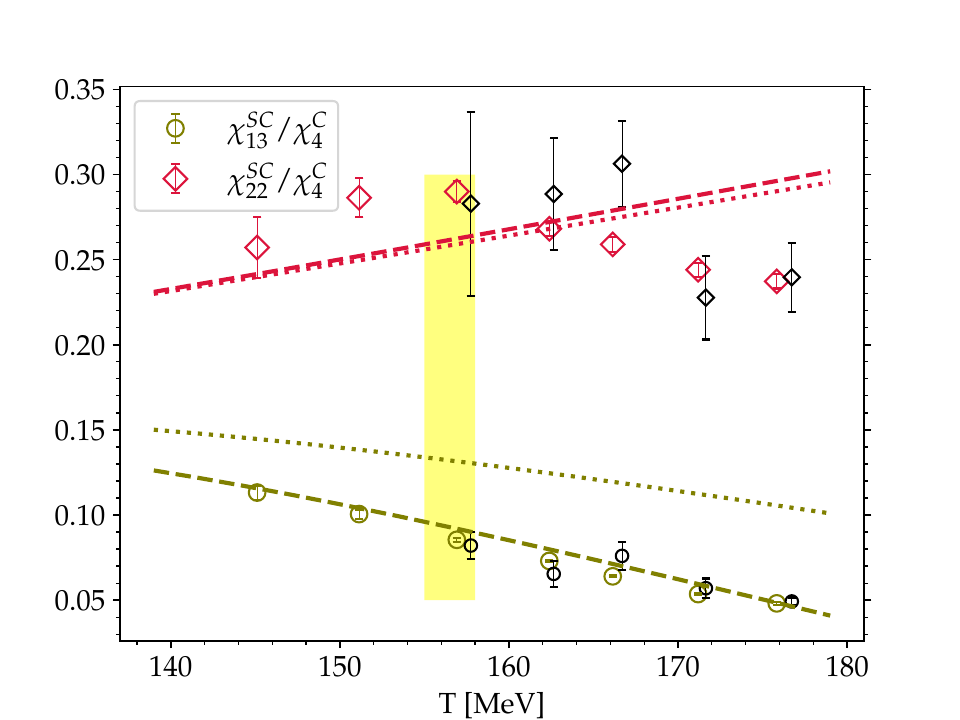}
\caption{Shown are the two $SC$ correlations normalised by $\chi^C_4$. The dashed (dotted) lines with the same color as the markers represent the respective QM-HRG (PDG-HRG) predictions. The black points with the same symbols as the $N_\tau=8_{[b]}$ results, represent the respective  $N_\tau=12_{[b]}$ data.}
\label{fig:sc_ratios}
\end{figure}
Finally let us discuss the ratios involving strangeness charm correlation.
In Fig. \ref{fig:sc_ratios} we show our results for $\chi^{SC}_{13}/\chi_4^C$ and $\chi^{SC}_{22}/\chi_4^C$.
The former ratio agrees well with the QM-HRG but not
with PDG-HRG, while the latter is about $10\%$ above the QM-HRG predictions in the vicinity of $T_{pc}$. 
As already argued in Section \ref{sec:hrg}
this latter ratio is sensitive to the spectrum of $|S|=2$ charmed hadrons.
The deviations of $\chi^{SC}_{22}/\chi_4^C$ from HRG can be due
to the imprecise or incomplete knowledge of $|S|=2$ baryon spectrum.
By using Eq. \eqref{eq:chi_HRG}, it can be easily deduced that unlike $|S|=1$ charmed hadrons, $|S|=2$ hadrons contribute with a higher absolute weight to $\chi^{SC}_{22}$ in comparison to $\chi^{SC}_{13}$:
\begin{eqnarray}
 \chi^{SC}_{22}&=& P^{C,S=1}_{M}+P^{C,S=1}_{B}+4P^{C,S=2}_{B}\; , \nonumber \\
\chi^{SC}_{13}&=& P^{C,S=1}_{M}-P^{C,S=1}_{B}-2P^{C,S=2}_{B} \; ,
\label{eq:sc1322}
\end{eqnarray}
where $P_M^{C,S=1}$ is the partial pressure for strange charmed mesons, $P_B^{C,S=1}$ is the partial pressure for strange charmed baryons 
with strangeness one, and $P_M^{C,S=2}$ is the partial pressure of strange charmed baryons with
strangeness two.  The deviations observed between QM-HRG calculations and our data for the
ratio $\chi^{SC}_{22}/\chi_4^C$ may be due to
missing $|S|=2$ charmed baryons not even calculated in quark models models so far. 
Moreover, in our QM-HRG list we use masses obtained
in the relativistic quark model calculations \cite{Ebert:2011kk}. Non-relativistic quark models \cite{Roberts:2007ni,Yoshida:2015tia}, however, often give predictions for
the masses of strange charmed baryons that are up to 100 MeV lower than the prediction of the relativistic
quark model. Therefore, we also estimated the ratio $\chi^{SC}_{22}/\chi_4^C$ using a QM-HRG
model, where all the missing $|S|=2$ charmed baryon masses are shifted down by 100 MeV. We find
that this version of QM-HRG shows much smaller deviations from the lattice QCD results for
$\chi^{SC}_{22}/\chi_4^C$ but still agrees with the lattice QCD results for $\chi^{SC}_{13}/\chi_4^C$.
Lastly, we also see that the $N_{\tau}=8_{[b]}$ and $N_{\tau}=12_{[b]}$ data for these ratios agree
within errors.

To summarize this section: We see that except for $\chi^{SC}_{22}/\chi_4^C$, various ratios of generalized susceptibilities below $T_{pc}$
agree well with the corresponding values obtained in QM-HRG and are significantly above the ones
obtained in PDG-HRG. Thus our lattice QCD results provide clear evidence for missing charmed hadrons, especially
charmed baryons. The QM-HRG underpredicts $\chi^{SC}_{22}/\chi_4^C$, possibly indicating uncertainties in the  spectrum of the doubly-strange charmed hadrons in the current QM-HRG list. While this is not always evident, the QM-HRG description breaks down just above $T_{pc}$
signaling the appearance of new degrees of freedom. Finally, the cutoff effects largely cancel out in the 
ratios of the generalized susceptibilities.

\section{Partial charm pressures and the nature of charm degrees
of freedom}
\label{sec:partial}
\subsection{Quasi-particle description}
The breakdown of QM-HRG for several quantities above $T_{pc}$ signals the possible appearence of new degrees of freedom.
In Ref. \cite{Mukherjee:2015mxc} a quasi-particle model has been proposed in which the charm pressure is written 
in terms of the partial pressures of non-interacting charm quarks, charmed mesons and charmed baryons:
\begin{equation}
{P_C(T,\vec{\mu})=P_M^C(T,\vec{\mu})+P_B^C(T,\vec{\mu})+P_q^C(T,\vec{\mu})} \text{ .}
\label{eq:quasi}
\end{equation}
The above model simply extends the HRG relation introduced in Eq.~\eqref{eq:P} to incorporate charm quark pressure given by Eq.~\eqref{eq:pq}. In this
model as well, $\chi_4^C$ is a proxy for the total charm pressure. The proxies of charm quark, charmed meson and charmed baryon partial pressures at $\vec{\mu}=0$ can be expressed
in terms of the generalized susceptibilities as \cite{Mukherjee:2015mxc}:
\begin{align}
	P_{q}^{C}&=9(\chi^{BC}_{13}-\chi^{BC}_{22})/2\; , 
	\label{eq:partial-quasiq}\\
	P_{B}^{C}&=(3\chi^{BC}_{22}-\chi^{BC}_{13})/2\; , 
	\label{eq:partial-quasiB}\\
	P_{M}^{C}&=\chi^{C}_{4}+3\chi^{BC}_{22}-4\chi^{BC}_{13} \; .
	\label{eq:partial-quasiM}
\end{align}

Eq. (\ref{eq:partial-quasiq}) is not the only way to define the pressure from charm quark quasi-particles.
As already pointed out, in our previous work \cite{Bazavov:2023xzm},  independent constructions of operators that project onto observables with quantum numbers of
charm quarks were discussed. We showed that three such operators projecting onto states with quantum numbers i) $|B|=1/3$, $|C|=1$, ii) $|Q|=2/3$, $|C|=1$, iii) $|B|=1/3$, $|Q|=2/3$, $|C|=1$, vanish below $T_{pc}$, and give
consistent results at temperatures above $T_{pc}$. This consistency can only be achieved if the proposed quasi-particle describes the charm thermodynamics in the explored temperature range. Such an approach has been used very successfully
to define operators for various ordinary hadron quantum number channels. Here we use first operator with $|B|=1/3$, $|C|=1$, given in Eq.~\eqref{eq:partial-quasiq}, to extend the existing HRG model \eqref{eq:P} to quasi-particle model in Eq.~\eqref{eq:quasi}. 

The high statistics $N_{\tau}=8_{[b]}$ results of charm quark, charmed meson
and charmed baryon partial pressures normalised by $\chi_4^C$ has been accurately determined in \cite{Bazavov:2023xzm}. 
It also has been
shown that the charm quark partial pressure below $T_{pc}$ is zero within errors, whereas the charmed baryon and meson partial pressures remain to be significant above $T_{pc}$ \cite{Bazavov:2023xzm}.
The continuum limit of $\chi^C_4$ obtained in Sec.~\ref{sec:continuum}, enables us to convert fractional contributions of coexisting charm quarks and charmed hadrons to the total charm pressure, calculated in \cite{Bazavov:2023xzm}, to absolute contributions. 

\begin{figure}
 \includegraphics[width=0.45\textwidth]{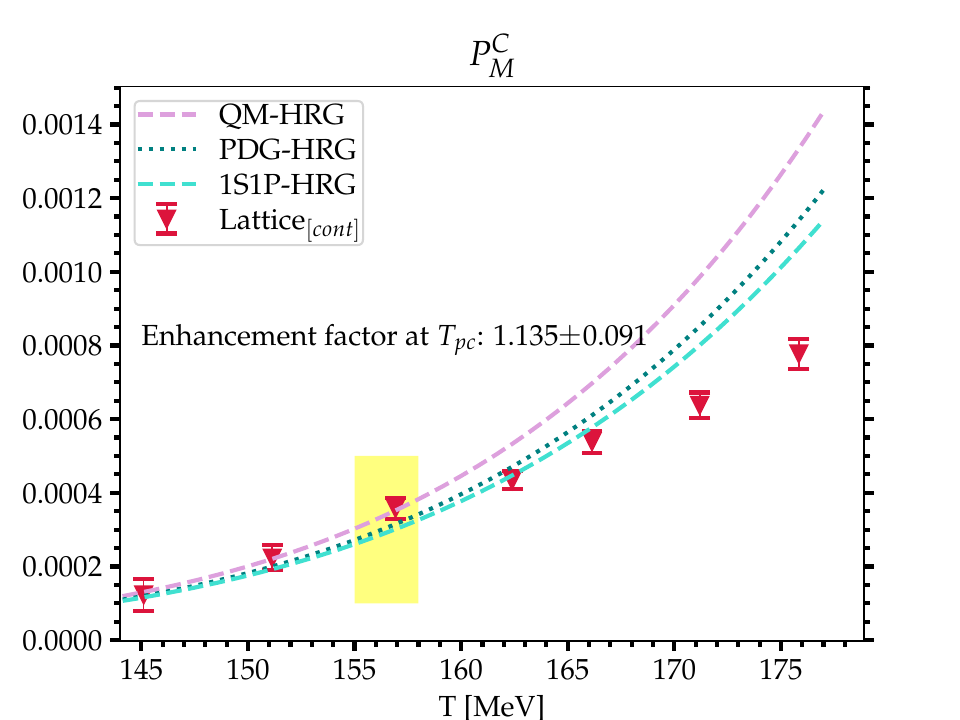}
 \includegraphics[width=0.45\textwidth]{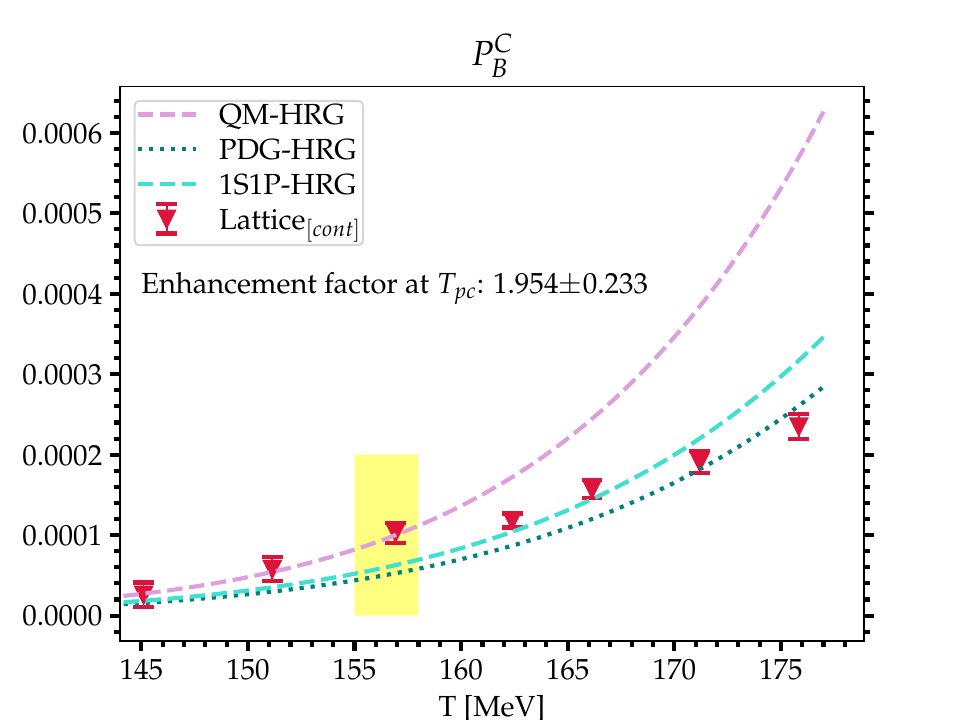}
\caption{Shown are the continuum estimates of the charmed meson [{\it top}] and charmed baryon [{\it bottom}] partial pressures as function of the temperature across
the chiral crossover. We compare the lattice QCD results with various versions of hadron resonance gas models. Enhancement factors are ratios of shown lattice quantities to their respective PDG-HRG counterparts. Note that after adding a few missing isospin partners to the PDG-HRG list, these factors slightly changed in comparison to our previous publication \cite{Sharma:2025zhe}.
}
\label{fig:pBM}
\end{figure}

In Fig. \ref{fig:pBM} we show the continuum estimates of the partial charmed meson and baryon pressures obtained in the quasi-particle model and compare to various HRG model calculations.
As expected, QM-HRG describes the partial pressures of charmed mesons and baryons up to
the crossover temperature, and PDG-HRG clearly underestimates both pressures. In particular, at $T_{pc}$, the lattice-QCD  result of $P_B^C$ is 
$1.95(23)$ times the PDG-HRG prediction of $P_B^C$, which implies that the partial charmed baryon pressure receives $51(6)\%$ contribution from experimentally known charmed baryons, while the rest comes from not-yet-discovered charmed baryons.  On the other hand, such an enhancement factor for $P_M^C$ is 
$1.13(9)$, which implies that at the chiral crossover, the partial charmed meson pressure receives $88(7)\%$ contribution from experimentally known charmed mesons. Therefore, as expected, charmed meson sector is more complete in comparison to the charmed baryon sector. These conclusions align with the prediction shown in Fig.~\ref{fig:barCmesC}. 
\begin{figure}
\includegraphics[width=0.45\textwidth]{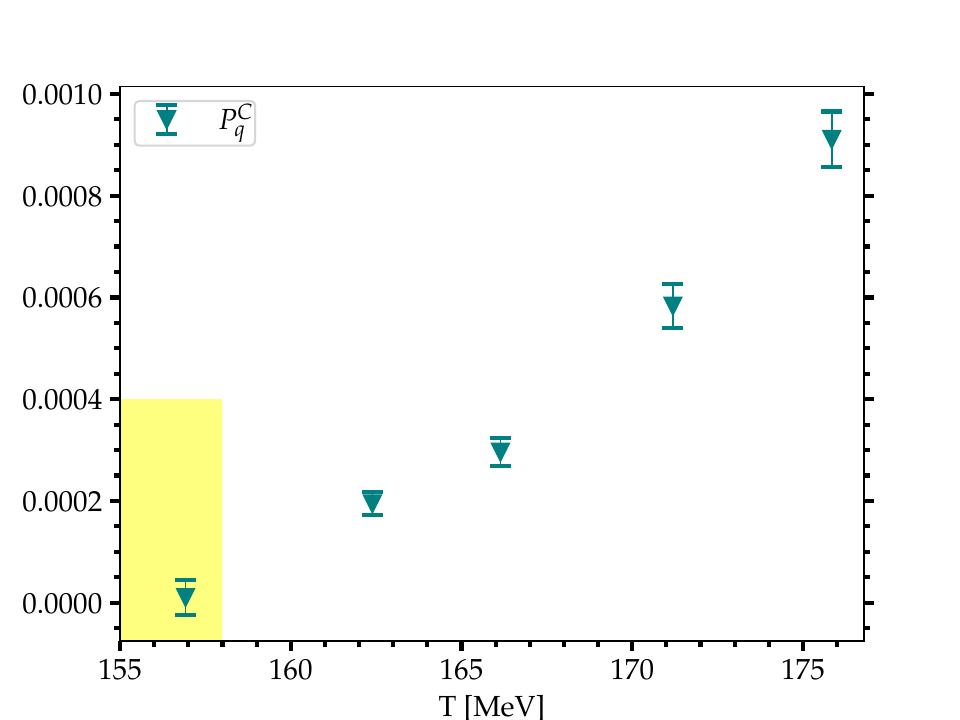}
\includegraphics[width=0.45\textwidth]{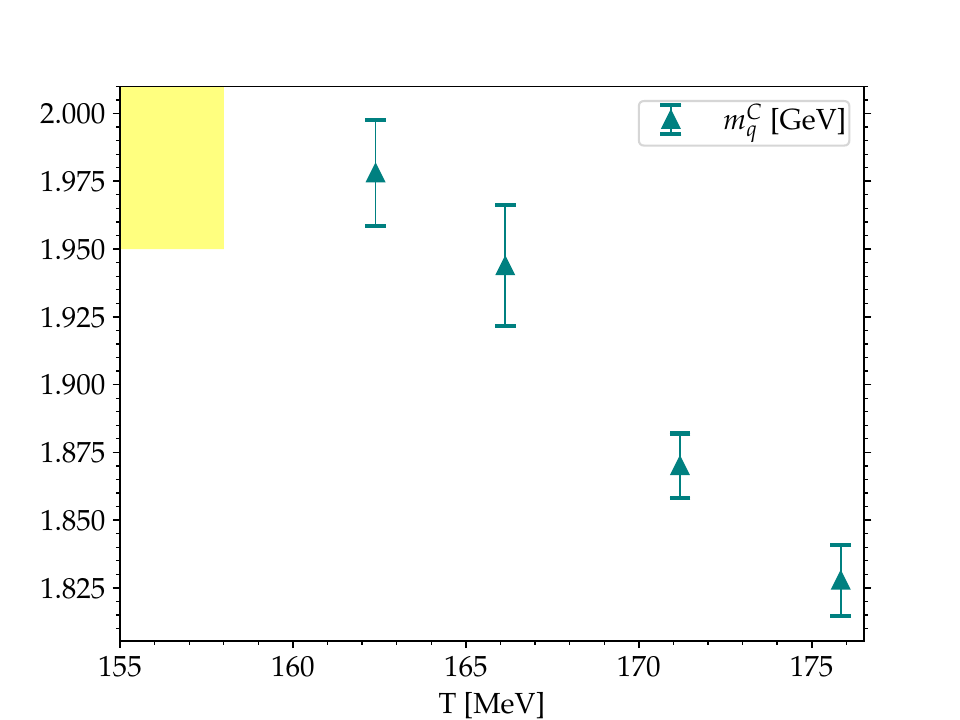}
	\caption{Shown are the continuum estimates of the partial pressure of charm quark-like excitation above $T_{pc}$ [{\it top}] and the temperature dependent in-medium mass of a charm quark-like excitation above $T_{pc}$ [{\it bottom}]. The yellow bands represent $T_{pc}$ with its uncertainty. Note that $m_q^C$ differs from \cite{Sharma:2025zhe}, where we only considered a gas of charm quarks and not quarks-antiquarks.}
	\label{fig:Pq}
\end{figure}

Above the crossover temperature, QM-HRG over-predicts the lattice results on charmed mesons and baryon pressures. This may imply that many of the charmed hadron states melt above the crossover temperature.  To explore this possibility a bit further we constructed a hadron resonance gas model which only contains the ground state charmed mesons and 
baryons and their parity partners, i.e. the first orbital excitation of these hadrons with vacuum masses. In the language of quark model, this means that we only consider the 1S and 1P mesons and baryons given in Tab. I, II and VII of \cite{Chen:2022asf}.
Therefore, we call this model the 1S1P-HRG model. The 1S1P-HRG model is closer to the lattice
results on the charmed mesons and charmed baryons pressures above the crossover temperature compared
to the QM-HRG model. 
We note, however, that lattice QCD calculations of charmed meson and charmed baryon correlation functions indicate
that around and above the crossover temperature in-medium modication of charmed hadrons masses could be
significant \cite{Aarts:2022krz,Aarts:2023nax}. In particular is was shown that the single charmed baryon masses
are sensitive to the chiral crossover \cite{Aarts:2023nax}. Therefore, the 1S1P-HRG model can only serve as a rough guidance
on charmed meson and charmed baryon pressure above $T_{pc}$.

In the coexistence phase of charmed hadron and quark-like excitations, $m^C_q$ in Eq. \eqref{eq:pq} becomes temperature dependent and can be interpreted as the mass of a quasi-particle with quantum numbers of charm quark. Given $P^C_q(T)$ as a function of temperature, one can solve Eq. \eqref{eq:pq} at each temperature and obtain temperature dependence of $m^C_q$. In Fig. \ref{fig:Pq} [{\it bottom}], at each temperature, the error on $m^C_q$ is the standard deviation of $m^C_q$ values obtained after solving Eq.~\eqref{eq:pq} for $50$ fake Gaussian samples. Fig.~\ref{fig:Pq} [{\it bottom}] shows that at $T=162.4$ MeV, $m^C_q$ is around the mass of D-meson, and starts decreasing with temperature. Quasi-particle model in Eq.~\eqref{eq:quasi} considers a non-interacting gas of charmed hadron and quark-like excitations, but the temperature dependence of $m^C_q$ encodes the in-medium interactions. In the non-interacting quark gas limit, $m^C_q$ will become the pole mass of charm quark.
\begin{figure*}
 \includegraphics[width=0.45\textwidth]{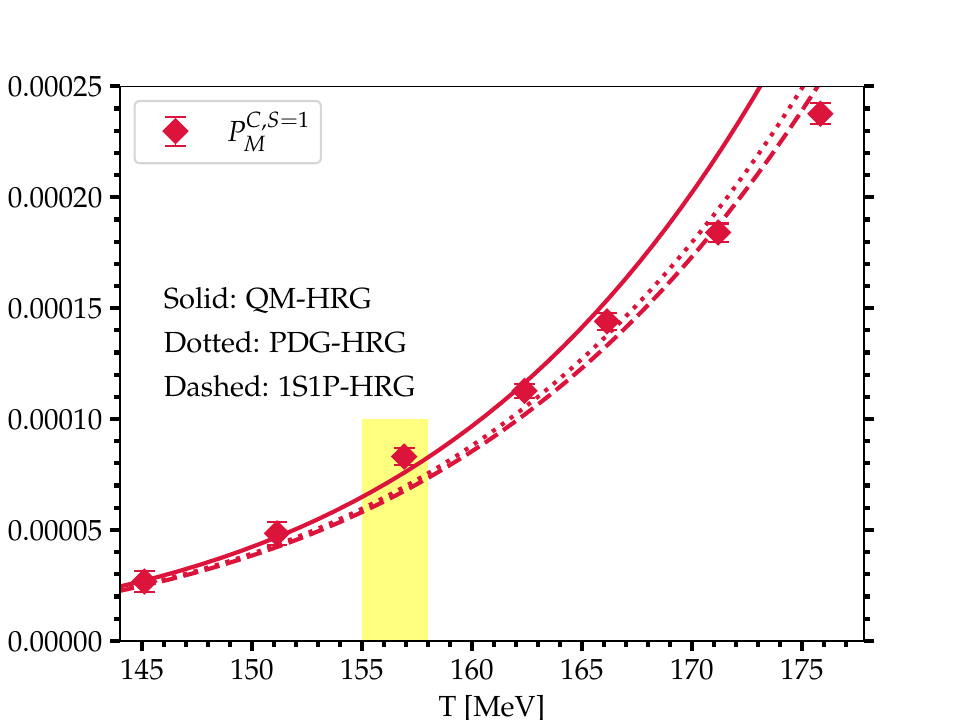}
 \includegraphics[width=0.45\textwidth]{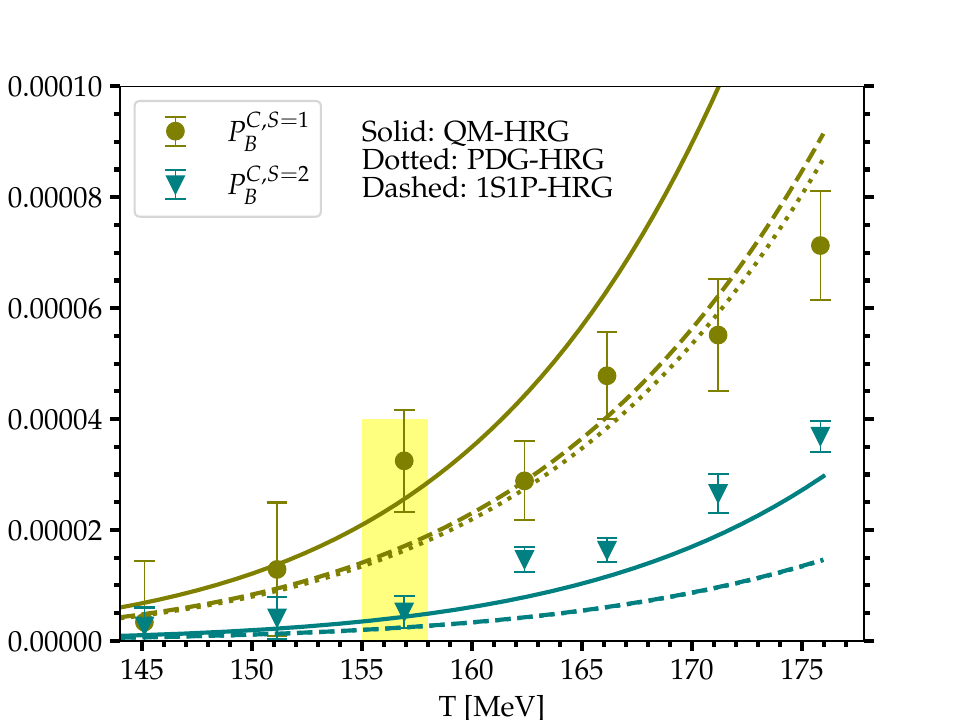}
\caption{Shown are the continuum  estimates of the strange charmed meson ({\it left}) and strange charmed baryon ({\it right}) pressures as function of the temperature across
the chiral transition. We compare the lattice QCD results with various versions of hadron resonance
gas model. Dashed: QM-HRG, Dotted: PDG-HRG, Solid: 1S1P-HRG.}
\label{fig:pSC}
\end{figure*}
\subsection{Strange charmed degrees of freedom}
\begin{figure}[t]
\includegraphics[width=\linewidth]{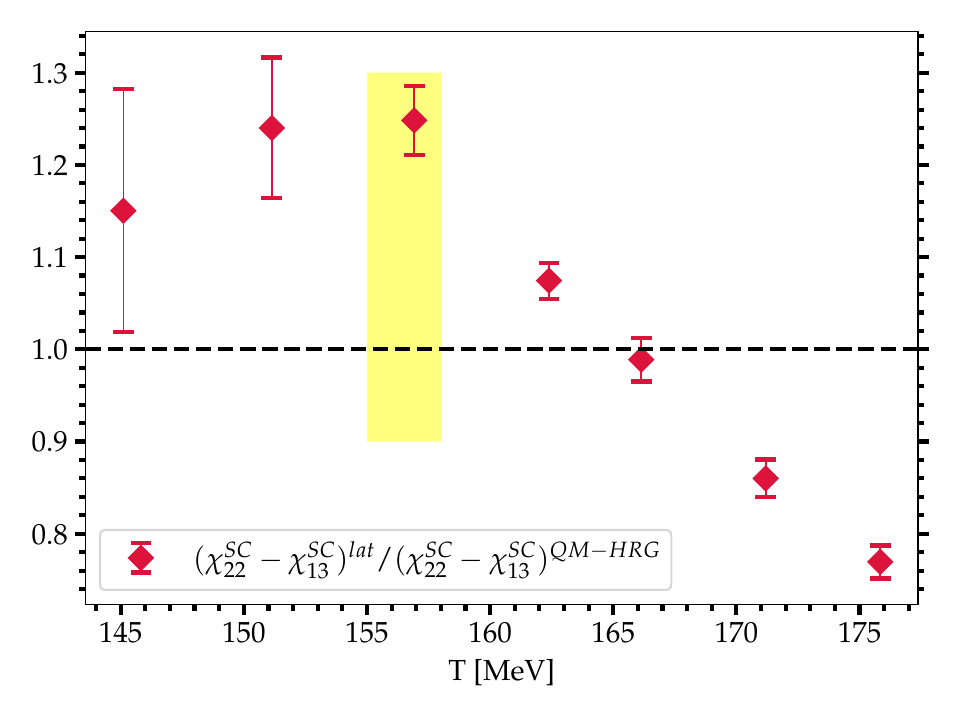}
\caption{Shown is the continuum estimate of the proxy for $2P_B^{C,S=1}+6P_B^{C,S=2}$ normalised by its QM-HRG expectation.}
\label{fig:norm}
\end{figure}
The charmed meson and charmed baryon pressures can be further decomposed into the partial pressures of mesons and baryons corresponding to different strangeness content. Using lattice QCD results on
the generalized susceptibilities involving strangeness, $\chi_{kmn}^{BSC}$ one can construct $P_M^{C,S=1}$, $P_B^{C,S=1}$ and $P_M^{C,S=2}$. The following explicit formulas for $P_M^{C,S=1}$, $P_B^{C,S=1}$ and $P_M^{C,S=2}$ in terms of $\chi_{kmn}^{BSC}$ are taken from Ref. \cite{Mukherjee:2015mxc}:
\begin{eqnarray}
P_{M}^{C,S=1}&=&\chi_{13}^{SC}-\chi_{112}^{BSC}\\
P_{B}^{C,S=1}&=&\chi_{13}^{SC}-\chi_{22}^{SC}-3\chi_{112}^{BSC}\\
P_{B}^{C,S=2}&=&(2\chi_{112}^{BSC}+\chi_{22}^{SC}-\chi_{13}^{SC})/2
\end{eqnarray}
\begin{figure}[t]
\includegraphics[width=\linewidth]{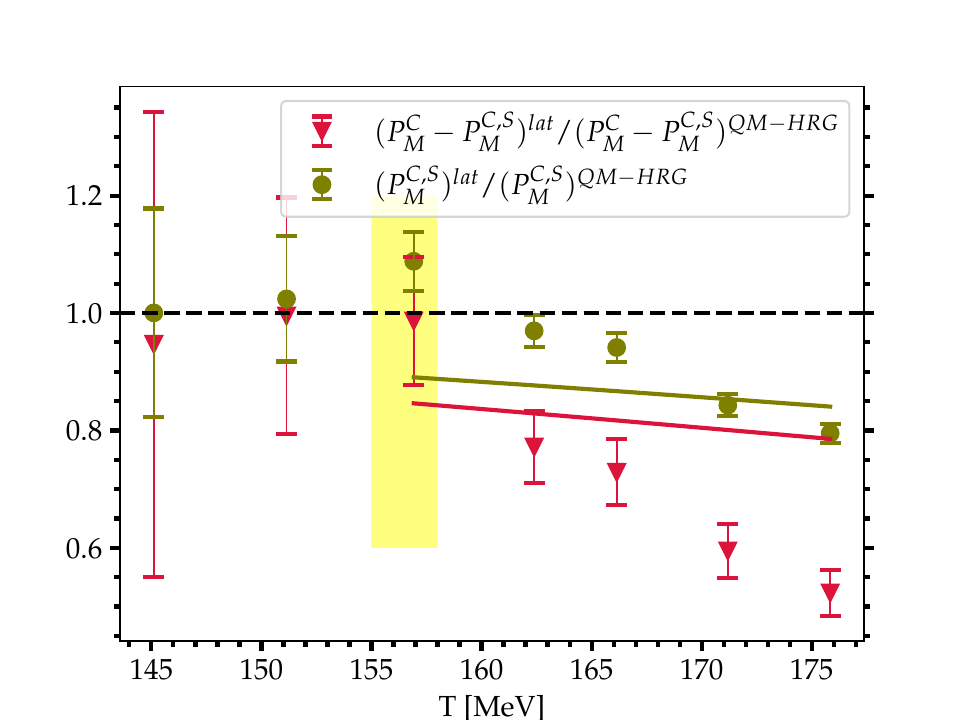}
\caption{Shown are the continuum estimates of the lattice results for non-strange and strange charmed meson pressures normalised by their corresponding QM-HRG counterparts. The solid lines show 1S1P-HRG predictions normalised by the corresponding QM-HRG predictions.}
\label{fig:mesons}
\end{figure}
\begin{figure}[t]
\includegraphics[width=\linewidth]{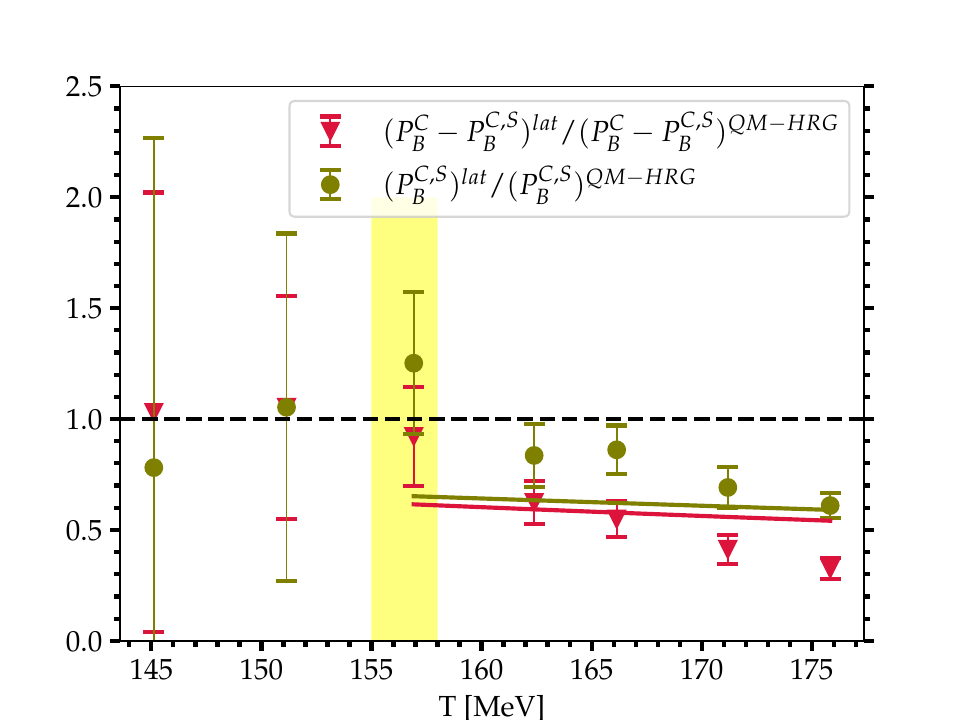}
\caption{Shown are the continuum estimates of the lattice results for non-strange and strange charmed baryon pressures normalised by their corresponding QM-HRG counterparts. The solid lines show 1S1P-HRG predictions normalised by the corresponding QM-HRG predictions.}
\label{fig:baryons}
\end{figure}
Our continuum estimates for $P_M^{C,S=1}$ ($P_M^{C,S}$), $P_B^{C,S=1}$ and $P_M^{C,S=2}$ are shown in Fig. \ref{fig:pSC}. The corresponding partial pressures normalised by $\chi^C_4$ for $N_\tau=8_{[b]}$ were shown in \cite{Sharma:2024edf}. 
We compare continuum estimates for these partial pressures with QM-HRG, PDG-HRG and 1S1P-HRG. These non-vanishing  strangeness-charm correlations above $T_{pc}$ support the existence of charmed hadrons in QGP. 
On the other hand the existence of strange charmed diquarks is not supported by the lattice results on generalized charm
susceptibilities \cite{Sharma:2024edf}.
As shown in Fig.~\ref{fig:pSC} ({\it left}), QM-HRG description works well for $P_M^{C,S}$ at low temperatures and breaks down above the
crossover temperature. On the other hand, given the current errors, it is difficult to establish disagreement with PDG-HRG and 1S1P-HRG in the strange charmed baryon sector, see Fig.~\ref{fig:pSC} [{\it right}]. Above $T_{pc}$, QM-HRG fails to describe the lattice results for strange charmed hadrons. Interestingly, for $|S|=2$ baryons, QM-HRG slightly under-predicts the lattice results above $T_{pc}$. This again may signal an inadequate knowledge of the $|S|=2$ charmed baryon sector that is used in our QM-HRG list. To explicitly see this, we analyse the continuum estimate of a lesser noisy observable which receives contributions only from the strange charmed baryon sector:
\begin{equation}
\chi_{22}^{SC}-\chi_{13}^{SC}=2P_B^{C,S=1}+6P_B^{C,S=2}\; .
\label{eq:proxy}
\end{equation}
Fig.~\ref{fig:norm} shows continuum estimate of \eqref{eq:proxy} normalised by its QM-HRG counterpart. At $T_{pc}$, the central value of continuum result differs by 25\% from the QM-HRG prediction. 
 If we then lower the masses of the missing $|S|=2$ charmed baryons by $100$ MeV we find that the continuum estimate for this quantity,
normalized by the corresponding QM-HRG result, is much closer to unity.

Having estimated the pressure of strange charmed hadrons we can also estimate the partial pressures of non-strange charmed hadrons, e.g.
$P_M^C-P_M^{C,S}$ is the pressure of non-strange charmed mesons. In Fig.~\ref{fig:mesons}, we show the ratios of the partial pressures of strange and non-strange charmed mesons to their respective counterparts calculated in the QM-HRG model. These ratios equal unity within errors below $T_{pc}$ as expected from Figs. \ref{fig:pBM} ({\it top}) and \ref{fig:pSC} ({\it left}). 
Above $T_{pc}$,
these ratios are below one. However, we see that the deviations from unity are significantly larger 
for non-strange charmed mesons than for strange charmed mesons. As can be seen, 1S1P-HRG can describe continuum estimate of non-strange charmed meson partial pressure for temperatures slightly above $T_{pc}$, whereas for strange charmed meson partial pressure, it works in a small temperature range above $170$ MeV. In Fig. \ref{fig:baryons} we show similar ratios for strange, $P^{C,S}_{B}=P^{C,S=1}_{B}+P^{C,S=2}_{B}$, and non-strange, $P_B^C-P^{C,S}_{B}$, charmed baryons and
again we see larger deviations from unity for non-strange charmed baryons. This may imply that more strange charmed hadrons can survive above the QCD crossover than non-strange
charmed hadrons. In other words non-strange charmed hadrons are more susceptible to the chiral crossover.  This is a reflection of QM-HRG underpredicting the partial pressures of strange charmed baryons due to additional missing states. In Figs. \ref{fig:mesons} and \ref{fig:baryons} we
also show the partial pressures of strange and 
non-strange charmed hadrons in 1S1P-HRG as solid lines. We see that at the highest two temperatures the 1S1P-HRG lies closer to the lattice data for strange hadron
pressures than for non-strange hadron pressures, i.e. 1S1P-HRG captures the lattice results for the strange hadron pressure better at the highest two temperatures.
However, if one could account for missing strange charmed hadrons in 1S1P-HRG, solid lines might be able to describe both strange and non-strange charmed hadron partials pressures in the small temperature range above $T_{pc}$.

\section{Conclusions}

In this paper we presented a comprehensive study of generalized charm susceptibilities at three
values of the lattice spacing with the aim to gain insights into the change of charm degrees of
freedom across the chiral crossover. It turned out that a large part of the cutoff effects is
associated with the choice of the charm quark mass in the region of the 
lattice spacings used in our study.
We showed that by setting the charm quark mass through the $D$-meson mass  one minimizes the cutoff
effects in generalized charm susceptibilities. Furthermore, we found that the ratios of generalized
susceptibilities are largely independent of the lattice spacing and the precise value of the charm
quark mass. Based of these observations we could provide continuum estimates of the generalized
charm susceptibilities. 

Above $T_{pc}$ lattice QCD results on generalized charm susceptibilities can be well described by a hadron
resonance gas model, which includes the so-called missing hadron states predicted by quark
model calculations, the QM-HRG. In a comprehensive study of various generalized charm 
susceptibilities we also showed that the QM-HRG model breaks down just above $T_{pc}$, even though
this breakdown is not apparent in some of the considered quantities, e.g. $\chi_4^C$.

To explain the temperature dependence of generalized susceptibilities above $T_{pc}$ we
used the quasi-particle model first proposed in Ref. \cite{Mukherjee:2015mxc}, in which the
charm pressure is written as sum of partial pressures for charmed mesons, charmed baryons and charm
quarks. From the charm quark pressure, which assumes a non-zero value above $T_{pc}$, we
extracted the quasi-particle mass of charm quarks. The partial charmed meson and 
baryon pressures drop below the corresponding QM-HRG values for $T>T_{pc}$, indicating that many higher lying
charmed hadron states melt above the chiral crossover. Within the quasi-particle model
of Ref. \cite{Mukherjee:2015mxc} we further decomposed the partial pressure of charmed
baryons and mesons into the partial pressures of strange and non-strange charmed baryons, and strange and non-strange charmed mesons. We found that the partial pressure of strange charmed hadrons show smaller deviations from QM-HRG above $T_{pc}$ 
compared to the partial pressures of non-strange charmed hadrons. This may indicate a different melting
pattern of strange and non-strange charmed hadrons. We also argued that $\chi_{22}^{SC}$ is very sensitive 
to both the number and the masses of missing strangeness two charmed baryons, and that the masses of these baryons could be up to $100$ MeV smaller
than predicted by quark model of Ref. \cite{Ebert:2011kk}.
\section*{Data availability}
All data from our calculations, presented in the figures
of this paper, can be found in Ref. \cite{datapublication}.

\acknowledgments
This work was supported by The Deutsche Forschungsgemeinschaft (DFG, German Research Foundation) - Project numbers 315477598-TRR 211 and 460248186 (PUNCH4NFDI). The authors gratefully acknowledge the computing time provided to them on the high-performance computer Noctua 2 at the NHR Center PC$^2$ under the project name: hpc-prf-cfpd. These are funded by the Federal Ministry of Education and Research and the state governments participating on the basis of the resolutions of the GWK for the national high-performance computing at universities (www.nhr-verein.de/unsere-partner). This work was also supported by the U.S. Department of Energy, Office of Science, Office of Nuclear Physics through Contract No. DE-SC0012704 and through
HEFTY Topical Collaboration in Nuclear Theory. We also acknowledge support by the DFG cluster of excellence ORIGINS funded by DFG under Germany's Excellence Strategy-EXC-2094-390783311 and the DFG Grant No. BR 4058/5-1.

\bibliography{refs}

\end{document}